\documentclass{article}          % twocolumn
\usepackage{graphicx}
\usepackage{subfigure}
\usepackage{color}
\newcommand{\grad}{\vec{\nabla}}

\newcommand{\lap}{\nabla^2}

\newcommand {\vu}     {{\bf u}}
\newcommand {\vx}     {{\bf x}}
\newcommand {\vr}     {{\bf r}}
\newcommand {\vxi}    {\mbox{\boldmath $\xi$}}
\newcommand {\CT}     {{\cal T}}

\begin{document}
%%%%%%%%%%%%%%%%%%%%%%%%%%%%%%%%%%%%%%%%%%%%%%%%%%%%%%%%%%%%%%%%%%%
%%%%%%%%%%%%%%%%%%%%%%%%%%%%%%%%%%%%%%%%%%%%%%%%%%%%%%%%%%%%%%%%%%%

\title{Mobility tensor of a sphere moving on a super-hydrophobic wall:
application to particle separation}
%%%%%%%%%%%%%%%%%%%%%%%%%%%%%%%%%%%%%%%%%%%%%%%%%%%%%%%%%%%%%%%%%%%
%%%%%%%%%%%%%%%%%%%%%%%%%%%%%%%%%%%%%%%%%%%%%%%%%%%%%%%%%%%%%%%%%%%


%\subtitle{Mobility tensor and dynamical system properties. \\ }


\author{D. Pimponi \and M. Chinappi \and P. Gualtieri \and C.M. Casciola}


\date{}

\maketitle

%%%%%%%%%%%%%%%%%%%%%%%%%%%%%%%%%%%%%%%%%%%%%%%%%%%%%%%%%%%%%%%%%%%
%%%%%%%%%%%%%%%%%%%%%%%%%%%%%%%%%%%%%%%%%%%%%%%%%%%%%%%%%%%%%%%%%%%
\begin{abstract}
The paper addresses the hydrodynamic behavior of a sphere close to a micro-patterned 
superhydrophobic surface described in terms of alternated no-slip and perfect-slip 
stripes. Physically, the perfect-slip stripes model the parallel grooves where a 
large gas cushion forms between fluid and solid wall, giving rise to slippage  at 
the gas-liquid interface. The potential of the boundary element method (BEM)  
in dealing with mixed no-slip/perfect-slip boundary conditions is exploited to 
systematically calculate the mobility tensor for different particle-to-wall  
relative positions and for different particle radii. The particle hydrodynamics  
is characterized by a non trivial mobility field which presents a distinct near 
wall behavior where the wall patterning directly affects the particle  motion. 
In the far field, the effects of the wall pattern can be accurately represented 
via an effective description in terms of a homogeneous wall with a suitably defined 
apparent slippage. The trajectory of the sphere under the action of an external 
force is also described in some detail.  
A ``resonant'' regime is found when the 
frequency of the transversal component of the force matches  a characteristic 
crossing frequency imposed by the wall pattern. It is found that, under resonance, 
the particle undergoes a mean transversal drift. Since the resonance condition 
depends on the particle radius the effect can in  principle be used to conceive 
devices for particle sorting based on superhydrophobic surfaces.

\end{abstract}
%%%%%%%%%%%%%%%%%%%%%%%%%%%%%%%%%%%%%%%%%%%%%%%%%%%%%%%%%%%%%%%%%%%
%%%%%%%%%%%%%%%%%%%%%%%%%%%%%%%%%%%%%%%%%%%%%%%%%%%%%%%%%%%%%%%%%%%

%%%%%%%%%%%%%%%%%%%%%%%%%%%%%%%%%%%%%%%%%%%%%%%%%%%%%%%%%%%%%%%%%%%
%%%%%%%%%%%%%%%%%%%%%%%%%%%%%%%%%%%%%%%%%%%%%%%%%%%%%%%%%%%%%%%%%%%
\section{Introduction} \label{sec:intro}

Super-hydrophobic (SH) surfaces have raised a large interest in the last 
decades for their 
self-cleaning~\cite{nosonovsky2009superhydrophobic,bottiglione2012role} and drag 
reducing~\cite{ybert2007achieving,ng2010apparent,lee2011influence,vinogradova2011wetting}
properties. These features are associated with gas or vapor bubbles trapped into 
the asperities of the solid surface. Commonly SH surfaces are fabricated by 
patterning the solid substrate with regular micro-structures (holes, grooves, 
pillars). In presence of a hydrophobic substrate, the liquid, usually water, 
hardly penetrates the hollows. This state is called the Cassie-Baxter state 
(also known as {\sl fakir}-state). In the Cassie-Baxter state the liquid is 
in contact with a patterned boundary consisting of alternated regions of 
liquid-solid and liquid-air/vapor interfaces. 

Since the trapped air or vapor acts as an almost perfect-slip cushion a simple 
model of a SH surface is given as a smooth wall with patterned boundary 
conditions. The standard no-slip condition applies to the liquid-solid interface 
and the perfect-slip condition at the liquid-air/vapor 
interface~\cite{ng2010apparent,philip1972flows}.

\begin{figure}
\includegraphics[width=0.9\textwidth]{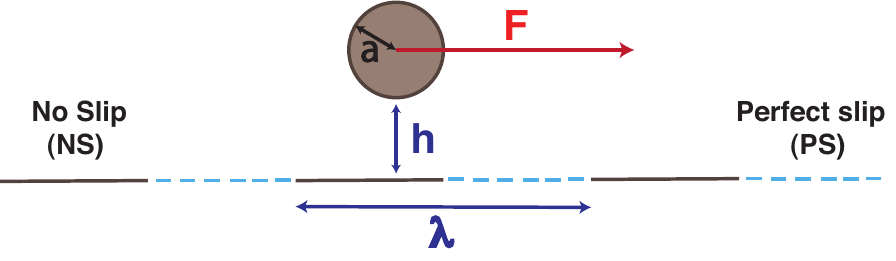}
\caption{System geometry. A sphere of radius $a$ moves 
at a distance $h$ from a planar wall. The wall is a superhydrophobic 
surface in Cassie state with a flat meniscus. The pattern
period is indicated with $\lambda$. The perfect-slip condition is 
used at the  air/liquid interface (PS, dashed line) and the no-slip condition is used 
at the solid-liquid interface (NS, solid line).
\label{fig:system}
}
\end{figure}

Purpose of the present paper is analyzing the hydrodynamics of a micrometer bead 
moving close to a SH surface. The patterned wall is modeled with 
alternating perfect-slip/no-slip parallel stripes (Fig. \ref{fig:system}).
All the typical length scales (particle radius, wall pattern length, and
particle-wall gap) considered here are on the order of micrometers, sufficiently 
large to describe the fluid as a continuum obeying the Navier-Stokes equations
\cite{li2009critical,chinappi2008mass,benzi2006mesoscopic,cottin2004dynamics,gentili2013water}
with no slippage at the solid wall 
\cite{chinappi2011tilting,chinappi2010intrinsic,huang2008water,zhang2012molecular,zhu2012reconciling,cottin2008nanohydrodynamics,pan2012role}. 
\textcolor{black}{
Our model corresponds to the real case of sufficiently deep grooves where
it can be safely assumed perfect-slip at the liquid-gas interface,
see e.g. the discussion reported 
in~\cite{vinogradova1995drainage,belyaev2010effective}.
}
\textcolor{black}{
At the same time the particle is sufficiently small to neglect 
fluid inertia in the limit of vanishing Reynolds number. 
Following standard dimensional analysis, the fluid acceleration in 
the Navier Stokes equations can be neglected  altogether leading 
to the linearized time independent Stokes equations \cite{happel1965low}.}
\textcolor{black}{
In these conditions, the coupling between the particle and the fluid
is entirely described by the mobility 
tensor field ${\vec M}({\vec x})$. 
In fact, for any particle position $\vec x$
the solution of the Stokes problem is achieved by using the 
so-called Boundary Element Method (BEM) where the system of partial 
differential equations is rewritten in terms of a vector boundary integral 
equation. The unknown reduces to the complementary data at the boundary, 
the stress vector where velocity (no-slip) is enforced or the velocity where 
the stress is prescribed (perfect-slip). From the Stokes solution the 
mobility tensor can be calculated, see e.g.~\cite{kim2005microhydrodynamics} 
for details. The $\vec x$ dependence of the mobility field is then recovered 
by placing the particle at different positions with respect to the patterned
wall.} 
The mobility tensor summarizes all the relevant hydrodynamic information needed to 
solve for the particle trajectory, once external forces are applied. All the 
complexity due to the patterned wall is lumped together in the mobility tensor 
field allowing for a simple parametric study of the particle response.

%\textcolor{red}{The evaluation of the resistance tensor requires six 
%independent solutions of the Stokes problem where a specific
%rigid body motions is imposed at once. By integration of the stress vector 
%at the particle surface the corresponding forces and the torques are evaluated. 
%The elements of the resistance tensor can be easily computed since they 
%linearly relate the calculated force/torque to the imposed velocity/angular 
%velocity at the particle surface. The resistance tensor is finally inverted to
%give the mobility tensor.}

%%%%%%%%%%%%%%%%%%%%%%%%%%%%%%%%%%%%%%%%%%%%%%%%%%%%%%%%%%%%%%%%%%%%%%%%%%%%
%%%%%%%%%%%%%%%%%%%%%%%%%%%%%%%%%%%%%%%%%%%%%%%%%%%%%%%%%%%%%%%%%%%%%%%%%%%%

\section{Mathematical model} \label{sec:num}
The linear Stokes system,
%%%%%%%%%%%%%%%%%%%%%%%%%%%%%%%%%%%%%%%%%%%%%%%%%%%%%%%%%%%%%%%%%%%%%%%%%%%%
\begin{eqnarray}
\nabla \cdot \vec{u} =0  & \\
\lap{\vec{u}} - \grad{p} = 0 & \; , 
\label{eq:stokes}
\end{eqnarray}
%%%%%%%%%%%%%%%%%%%%%%%%%%%%%%%%%%%%%%%%%%%%%%%%%%%%%%%%%%%%%%%%%%%%%%%%%%%%
for the velocity $\vu$ and the pressure $p$, here written in dimensionless form
with half the perfect-slip stripe width $w$ as reference length and
$\mu/(w \rho)$ and  $\mu^2/(\rho w^2)$ as velocity and pressure scale respectively,
is recast in terms of a boundary integral formulation
%%%%%%%%%%%%%%%%%%%%%%%%%%%%%%%%%%%%%%%%%%%%%%%%%%%%%%%%%%%%%%%%%%%%%%%%%%%%
\begin{eqnarray}
\label{eq:integral_eq}
E(\vxi) u_j(\vxi)& = &\frac{1}{8 \pi}
\oint_{\partial \Omega}  t_i(\vx) G_{ij}(\vx,\vxi) \, dS_{\vx}  - \nonumber \\
&& \frac{1}{8\pi} \oint_{\partial \Omega} 
u_i(\vx)  \CT_{ijk} (\vx,\vxi) n_k(\vx) \, dS_{\vx}.
\end{eqnarray}
%%%%%%%%%%%%%%%%%%%%%%%%%%%%%%%%%%%%%%%%%%%%%%%%%%%%%%%%%%%%%%%%%%%%%%%%%%%%
Here $\partial \Omega$ is the boundary of the flow domain $\Omega$, which consists 
of the particle surface and the patterned wall. $G_{ij}$ is the fundamental solution
(free-space Green's function or Stokeslet), $\CT_{ijk}$ the associated stress tensor, 
$u_i$ the i-th velocity component  and $t_i$ the traction vector, 
$t_i = -p \delta_{ij} + \partial u_i/\partial x_j + \partial u_j/\partial x_i$ with 
$\delta_{ij}$ the Kronecker delta.
The velocity at $\vxi$ is expressed as the convolution of the densities (velocity 
and stress vector) with the appropriate convolution kernels (Green's function 
tensor and associated stress). The coefficient $E$ equals $1$ inside the fluid 
domain and $1/2$ at regular boundary points, respectively. Collocating the 
representation at boundary points provides a boundary integral equation where 
the unknowns are the complementary data to those prescribed at the boundaries 
(i.e. the unknown is the velocity where the  traction vector is prescribed and 
the traction vector where the velocity is given). In the present case the data 
are the velocity where  no-slip holds and a combination of vanishing wall-normal 
velocity and tangential traction at the impermeable perfect slip regions. 
The integral equation can be solved numerically by the so-called 
boundary element method \cite{pozrikidis1992boundary,kim2005microhydrodynamics},  
a standard approach for linear systems of partial differential equation with 
constant coefficients (see the Appendix for a few more details). After the 
traction at the particle surface is evaluated, the hydrodynamic forces and 
torques are obtained by integration thus determining the resistance matrix and, by 
inversion, the mobility tensor.

In all the cases considered below, the solid fraction $0 \le \phi_s \le 1$ of the patterned 
wall, ratio of no-slip to total surface area, is $\phi_s =0.5$ (equal width for 
the perfect-slip and the no-slip stripes), such that the dimensionless pattern 
periodicity is $\lambda = 4$ (see sketch in Fig.~\ref{fig:system}).
\textcolor{black}{No difficulty is found to extend the numerical model to other solid fractions, provide resolution issues are treated with adequate care 
for the extreme cases (very small or very large $\phi_s$).}
In the following, unless otherwise explicitly stated, only the case of a 
no-slip particle will be addressed.  Again, given the flexibility of the approach, no substantial difficulty is 
encountered for different boundary conditions (perfect-slip or partial slip 
boundary conditions at the particle boundary). The SH wall coincides with the 
$Oxy$ plane of the reference system with the alternating 
perfect-slip (PS) and no-slip (NS) stripes parallel to the $y$ axis 
(see Fig.~\ref{fig:system}) being $z$ the wall-normal coordinate. 
Where convenient, the different components of vectors and tensors will be 
denoted by indices, e.g.  $x\equiv x_1$, $y\equiv x_2$, and $z\equiv x_3$. 
The dimensionless sphere radius is $a$ with $h$  the normalized gap between 
sphere and wall. 

In the context of Stokes flows, the response is linear with respect to the 
external force applied to the particle. It follows that the problem of 
determining the linear and angular velocity of the sphere, given forces and 
torques, can be conveniently formulated in terms of the mobility tensor  
$M_{\alpha\,\beta}$, $\alpha,\,\beta=1,\ldots 6$, 
see e.g. the classical textbook~\cite{kim2005microhydrodynamics}. Indeed the 
mobility tensor relates the generalized (linear and angular) velocities to the 
generalized forces (forces and torques), namely
%%%%%%%%%%%%%%%%%%%%%%%%%%%%%%%%%%%%%%%%%%%%%%%%%%%%%%%%%%%%%%%%%%%%%%%%%%%%
\begin{equation}
\left[
\begin{array}{c}
U_1 \\  U_2 \\   U_3 \\
\omega_1 \\  \omega_2 \\  \omega_3 \\
\end{array}
\right]
=
\left[
\begin{array}{cccccc}
M_{11} & M_{12} & M_{13} & M_{14} & M_{15} & M_{16} \\
M_{21} & M_{22} & M_{23} & M_{24} & M_{25} & M_{26} \\
M_{31} & M_{32} & M_{33} & M_{34} & M_{35} & M_{36} \\
M_{41} & M_{42} & M_{43} & M_{44} & M_{45} & M_{46} \\
M_{51} & M_{52} & M_{53} & M_{54} & M_{55} & M_{56} \\
M_{61} & M_{62} & M_{63} & M_{64} & M_{65} & M_{66} 
\end{array}
\right]
\left[
\begin{array}{c}
F_1 \\  F_2 \\  F_3 \\
T_1 \\  T_2 \\  T_3 \\
\end{array}
\right]
\label{eq:Mcompl}
\end{equation}
%%%%%%%%%%%%%%%%%%%%%%%%%%%%%%%%%%%%%%%%%%%%%%%%%%%%%%%%%%%%%%%%%%%%%%%%%%%%
where $U_i$, $i=1,\ldots3$, are the components of the particle center velocity 
$\vec{U}$, $\omega_i$ the components of the angular velocity 
\mbox{\boldmath{$\omega$}}, and $F_i$ and $T_i$  are  the components of force 
$\vec{F}$ and  torque $\vec{T}$, respectively. With this notation  
$M_{i,\,j+3}$ gives the coupling between the j-th  component of the torque and 
i-th component of the linear velocity of the sphere.
The reciprocity theorem guarantees the symmetry of the mobility tensor, 
$M_{\alpha \beta} = M_{\beta \alpha}$. In more compact notation 
eq.~(\ref{eq:Mcompl}) is rewritten as
%%%%%%%%%%%%%%%%%%%%%%%%%%%%%%%%%%%%%%%%%%%%%%%%%%%%%%%%%%%%%%%%%%%%%%%%%%%%
\begin{equation}
\label{eq:M}
\vec{\tilde U}=\vec{M} \cdot \vec{\tilde F}
\end{equation}
%%%%%%%%%%%%%%%%%%%%%%%%%%%%%%%%%%%%%%%%%%%%%%%%%%%%%%%%%%%%%%%%%%%%%%%%%%%%
where  $\vec{\tilde U}$ and  $\vec{\tilde F}$ are the generalized velocities and
forces and $\vec{M}$ is the mobility tensor. In the general case the mobility is a 
tensor field $\vec{M}(\vec{x})$ depending on the position of the sphere center. 
The symmetry of the problem induces a corresponding symmetry on the tensor field.
For instance, in free space the mobility tensor reduces to a diagonal matrix
for a spherical body, consistently with the absence of any coupling among the different degrees of freedom. 
A homogeneous wall breaks the translation symmetry of the sphere in the 
wall-normal direction and induces the coupling between wall-parallel 
translations in direction ${\vec {\hat e}}$ and wall parallel rotations with 
axis parallel to ${\vec {\hat z}} \times {\vec {\hat e}}$ where 
${\vec {\hat e}}$  is any wall parallel  unit vector and ${\vec {\hat z}}$ is 
the wall-normal unit vector pointing towards the fluid, respectively, see e.g. 
the review \cite{brady1988stokesian} and the 
textbook~\cite{kim2005microhydrodynamics}.

%%%%%%%%%%%%%%%%%%%%%%%%%%%%%%%%%%%%%%%%%%%%%%%%%%%%%%%%%%%%%%%%%%%%%%%%%%%%
\section{Sphere moving along a homogeneous wall} \label{sec:numerical}

The solution of the Stokes equations (\ref{eq:stokes}) in the geometry 
described in Fig.~\ref{fig:system} is obtained by an in-house 
boundary element code based on the BEMLIB  library
\footnote{The library is available under the terms of the GNU 
Lesser General Public License at \texttt{http://dehesa.freeshell.org/BEMLIB/} }
released by Pozrikidis~\cite{pozrikidis2002practical}.
This approach allows to tackle complex boundaries where the
complexity relies both on the geometrical configuration and on the
assigned boundary conditions, the alternation of PS/NS regions in the 
present case, Appendix~\ref{sec:BEM}.
When dealing with wall bounded Stokes flows, the effect of a single planar wall 
can in principle be included in the Green's function (wall Green function) 
thus avoiding the discretization of the wall itself, see e.g. \cite{blake1971note}.
However the use of a specialized Green's function becomes too cumbersome to deal  
with alternated PS/NS boundary conditions at the wall as required to model the 
present SH surface. It is more convenient to work with the free-space Green's 
function and use a boundary integral equation extending to particle and wall 
surfaces. Clearly, the infinite planar wall is truncated in numerics where it 
is modeled as a finite square of size $L \gg \lambda$. The appropriate truncation 
length $L$, in  the range of sphere-to-wall clearance $h$ considered here, is selected 
by comparison with the results of the wall Green function formulation 
(corresponding to an actually infinite wall) for the simple case of a no-slip wall.

%%%%%%%%%%%%%%%%%%%%%%%%%%%%%%%%%%%%%%%%%%%%%%%%%%%%%%%%%%%%%%%%%%%
%%%%%%%%%%%%%%%%%%%%%%%%%%%%%%%%%%%%%%%%%%%%%%%%%%%%%%%%%%%%%%%%%%%
\subsection{No-slip wall}

A first validation of the numerics concerns the sphere in free space. 
As for the homogeneous wall, the reference length $w$ introduced in the previous 
section is, strictly speaking, undefined. It  is fixed in this case by requiring  
the dimensionless sphere radius to be  $a=1$.
The numerically estimated resistance of the no-slip sphere in free space was 
checked to recover the well known Stokes results for rigid body translation in 
direction ${\vec {\hat e}}$ and rigid body rotation 
${\vec \omega }$, ${\vec D}= - 6 \pi a {\vec {\hat e}}$ and 
${\vec Q} = -8 \pi a {\vec \omega}$, respectively. 
It may be worthwhile calling the reader's attention on the fact that 
${\vec F}$ and ${\vec T}$ as defined in eq.~(\ref{eq:Mcompl}) are external forces 
applied to the sphere while ${\vec D}$ and ${\vec Q}$ denote the drag force 
and torque experienced by the sphere in the relative motion with respect to the 
fluid. In other words, for constant translation and rotation velocities, 
${\vec D} + {\vec F} = 0$ and ${\vec Q} + {\vec T} = 0$.
As a second  check the drag law  for the perfect-slip sphere moving in direction 
${\vec {\hat e}}$, ${\vec D}=- 4\pi a {\vec {\hat e}}$ and ${\vec Q} = 0$, was 
also reproduced, see~\cite{landau1987fluid}.

%%%%%%%%%%%%%%%%%%%%%%%%%%%%%%%%%%%%%%%%%%%%%%%%%%%%%%%%%%%%%%%%%%%%%%%%%%%%
\begin{figure}[t!]
\centerline{
\includegraphics[width=0.9\textwidth]{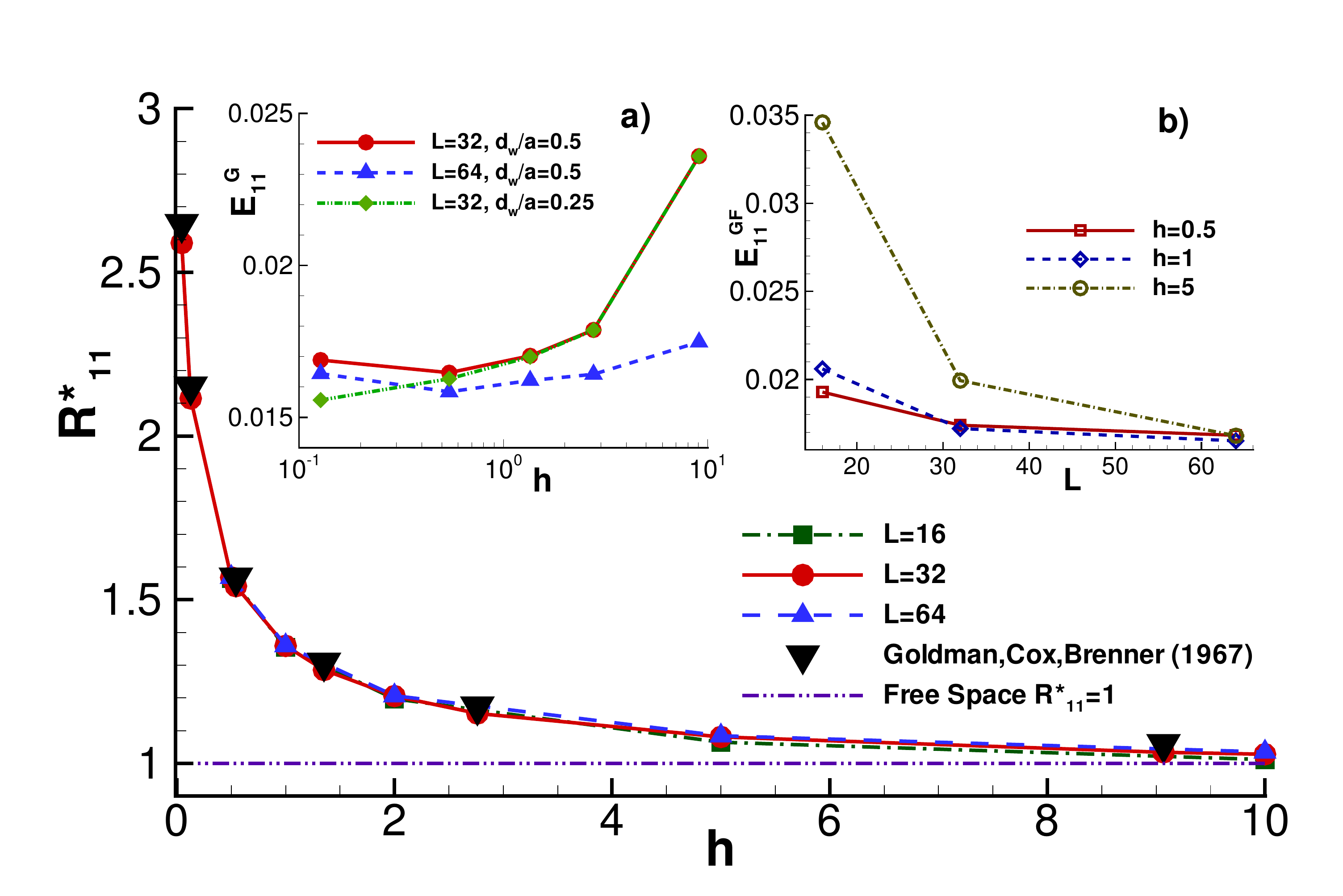}}
\caption{
\textcolor{black}{Main panel:
The $R_{11}^*$ resistance coefficient of a sphere of radius $a=1$ 
over a truncated no-slip wall plotted 
against the gap $h$ for three 
different truncation lengths: $L=16$ (squares, dash-dotted line), $L=32$ 
(circles, solid line), $L=64$ (triangles, dashed line). 
Black inverted triangles refer to 
the analytical solution by Goldman, Cox and Brenner~\cite{goldman1967slow}. 
Inset $a)$: relative error between the present data and~\cite{goldman1967slow}, 
$E^{G}_{11}=(R^{G}_{11}-R_{11})/R^{G}_{11}$ where the superscript $G$ refers to
Goldman et al., 
for $L=32$ and $L=64$.
Inset b): relative error 
$E^{GF}_{11}=(R^{GF}_{11}-R_{11})/R^{GF}_{11}$, where $GF$ refers 
to numerical results obtained by using the Green Function 
for an infinite no-slip wall~\cite{blake1971note},
plotted against $L$ for three different gaps: 
$h=0.5$ (squares, solid line), $h=1$ (diamonds, dashed line), 
$h=5$ (circles, dash-dotted line). }
\label{fig:flat}
}
\end{figure}
%%%%%%%%%%%%%%%%%%%%%%%%%%%%%%%%%%%%%%%%%%%%%%%%%%%%%%%%%%%%%%%%%%%%%%%%%%%%
More interesting are the tests in presence of the wall. In Stokes flows the 
velocity disturbance decays in space as the inverse distance from the momentum 
source, i.e. the sphere in the present case, as easily shown from the far field 
asymptotic of representation (\ref{eq:integral_eq}) where $G_{ij} \sim 1/r$. 
For this reason domain truncation effects must be carefully addressed.
In presence of a homogeneous wall, the hydrodynamic force ${\vec D}$ due to a 
wall-parallel translation of the sphere in direction ${\vec {\hat e}}$ has,
by symmetry, a vanishing wall normal component $D_3$. Indeed, $D_3$ depends linearly 
on the particle velocity and should change sign under velocity inversion. 
Clearly $D_3$  should instead be independent of the direction of the wall parallel 
velocity. The only possible conclusion is that $D_3 \equiv 0$. Hence the only 
non vanishing component of the hydrodynamic force is the one opposed to the 
velocity,
$$
R_{11} = R_{22} = -{\vec D} \cdot{\vec {\hat e}} = 
{\vec {\hat e}} \cdot \int_{\partial B} {\vec t} dS > 0 \, ,
$$
where $\vec t$ is the stress exerted by the body on the fluid.
\textcolor{black}{
Figure~\ref{fig:flat} shows the normalized resistance coefficient 
$R_{11}^* = R_{11}/(6 \pi a)$ for a no-slip 
sphere of radius $a = 1$ moving in the wall parallel direction ${\vec {\hat e}}_1$ close to a 
no-slip  wall. Data are reported as a function of the gap $h$ for different domain 
truncation lengths $L$. 
As $L$ is increased the results show apparent convergence toward the infinite wall 
result, compare data at $L=32$ and $L=64$. 
For further comparison the analytic results obtained in 
\cite{goldman1967slow} are also reported.}
\textcolor{black}{
The resistance coefficient $R^{GF}_{11}$ for the actual infinite planar wall was 
also obtained by using 
a companion numerical solution that employed the
wall Green function. 
The inset a) concerns the relative error, $E^{G}_{11}=(R^{G}_{11}-R_{11})/R^{G}_{11}$ where the superscript $G$ refers to
Goldman et al.,   between the present numerics and the analytical solution of \cite{goldman1967slow}.
The three curves correspond to a reference solution with $L = 32$ and a characteristic wall panel dimension 
$d_w/a = 0.5$, to a second solution with the 
same truncation length and a finer discretization $d_w/a = 0.25$, and to a third case with the same grid of the reference numerical solution and
increased truncation length $L = 64$. From the comparison it is apparent that the typical panel dimension controls the error when the particle is close to the wall.
Instead, when the wall-normal distance increases the effect of truncation becomes dominant, requiring a larger portion of the wall to be retained in the 
numerical configuration.
The inset b) reports the relative error  
$E^{GF}_{11} = (R_{11} - R^{GF}_{11})/ R^{GF}_{11}$ with respect to the wall-Green Function approach
vs the truncation length $L$ for different gaps $h$. As already commented on, $E^{GF}_{11}$ increases with $h$ at fixed wall truncation and 
decreases with $L$ at fixed wall distance, confirming that  a finite portion of the planar wall is seen to better approximate the 
infinite wall case when the distance of the object from the wall gets smaller and smaller.
For the typical gaps to be further considered in this paper, 
namely $h \in (0.125,2)$, no significant improvements are achieved by increasing 
$L$ from $32$ to $64$ with the relative error in both cases below $\sim 2\%$.  
Hence, where not explicitly stated, the value $L=32$ is used throughout the paper.  
Similar convergence is observed for all other non-zero  resistance tensor 
coefficients (data not shown).} 
\textcolor{black}{For this range of parameters the particle surface was
discretized by means of a hierarchical triangular mesh of
$512$ elements whose typical size is $d_p/a = 0.1567$. 
A non-uniform discretization consisting of about $1500$ elements is adopted for the wall.
In fact, the  tessellation of the wall is locally refined below the sphere 
and is progressively coarsened away from it.}
%%%%%%%%%%%%%%%%%%%%%%%%%%%%%%%%%%%%%%%%%%%%%%%%%%%%%%%%%%%%%%%%%%%%%%%%%%%%
\begin{figure}[t!]
\centerline{
\includegraphics[width=0.9\textwidth]{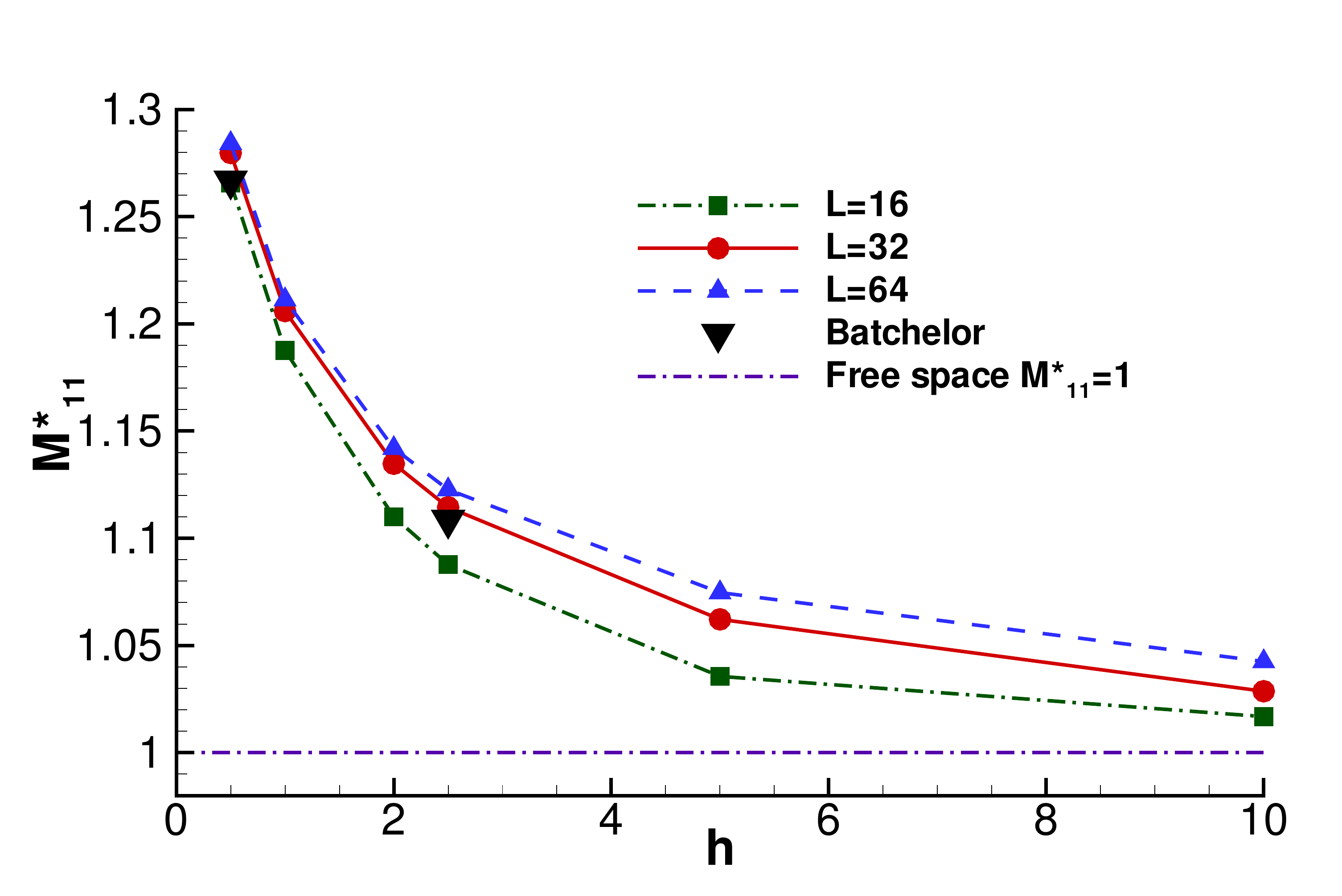}}
\caption{
\textcolor{black}{Mobility coefficient $M_{11}^*$ of a  no-slip solid sphere of 
radius $a=1$ in presence of a perfect-slip wall as a function of the 
sphere-to-wall gap $h$ for different  wall truncations $L=16$ (squares, dashed line),
$L=32$ (circles, dash-dotted line), $L=64$ (triangles, solid line). 
Black inverted triangles refer to the analytical solution by 
Batchelor \cite{batchelor1976brownian}.
}
\label{fig:flatFREE}
}
\end{figure}
%%%%%%%%%%%%%%%%%%%%%%%%%%%%%%%%%%%%%%%%%%%%%%%%%%%%%%%%%%%%%%%%%%%%%%%%%%%%

%%%%%%%%%%%%%%%%%%%%%%%%%%%%%%%%%%%%%%%%%%%%%%%%%%%%%%%%%%%%%%%%%%%
%%%%%%%%%%%%%%%%%%%%%%%%%%%%%%%%%%%%%%%%%%%%%%%%%%%%%%%%%%%%%%%%%%%
\subsection{Perfect-slip wall}

After the preliminary validation provided  in the previous subsection, numerical results worth being discussed concern the motion of a no-slip sphere 
close to  a perfect-slip homogeneous wall. 
\textcolor{black}{In this case symmetry considerations can be exploited to provide a reference solution to compare the present result with.
Indeed considering the image of the sphere with respect to the perfect slip wall one ends up with a two-sphere system translating parallel to the wall in otherwise 
infinite space. By symmetry it  is clear that taking the two-sphere solution restricted to the half-space above the wall provides the required solution for the present perfect-slip wall problem. The solution of two-sphere problem was discussed by Batchelor \cite{batchelor1976brownian} where reference data for the mobility 
are provided. Alternatively, one can exploit the equivalent two sphere problem to derive a boundary integral equation only involving the unknown traction on the  physical sphere that accounts for its image below the wall in such a way that the solution provides the desired results for the perfect slip wall case.
}

The main results are summarized in Fig.~\ref{fig:flatFREE} where the mobility 
coefficient $M_{11}^*$ for the sphere is reported as a function of 
the dimensionless gap $h$. Even in this case a few wall truncation lengths are 
considered, namely $L=16, \, 32, \, 64$. As expected $M_{11}^*$ 
decreases (resistance increases) approaching the 
free-space value for increasing gap. On the contrary the mobility coefficient 
increases (resistance lowers) when the gap is progressively reduced. This is a consequence of the 
perfect-slip boundary condition that allows a finite fluid velocity at 
the wall.  It follows that strong velocity gradients can not occur in the 
gap between sphere and wall in contrast to the case of a no-slip surface.  
\textcolor{black}{Also in this case the numerical solution shows good agreement with respect to the equivalent Batchelor 
two-sphere problem (black inverted triangles in Fig.~\ref{fig:flatFREE}).}
Even for the perfect-slip wall the results are only negligibly affected by the wall 
truncation for sufficiently large  $L$ , as shown by comparing $M_{11}^*$ at $L=32$ 
and $L=64$ for the present values of the gap $h\in(0.125,2)$.
%%%%%%%%%%%%%%%%%%%%%%%%%%%%%%%%%%%%%%%%%%%%%%%%%%%%%%%%%%%%%%%%%%%%%%%%%%%%
\begin{figure*}
\subfigure[a=0.25]{
\includegraphics[width=0.5\textwidth]{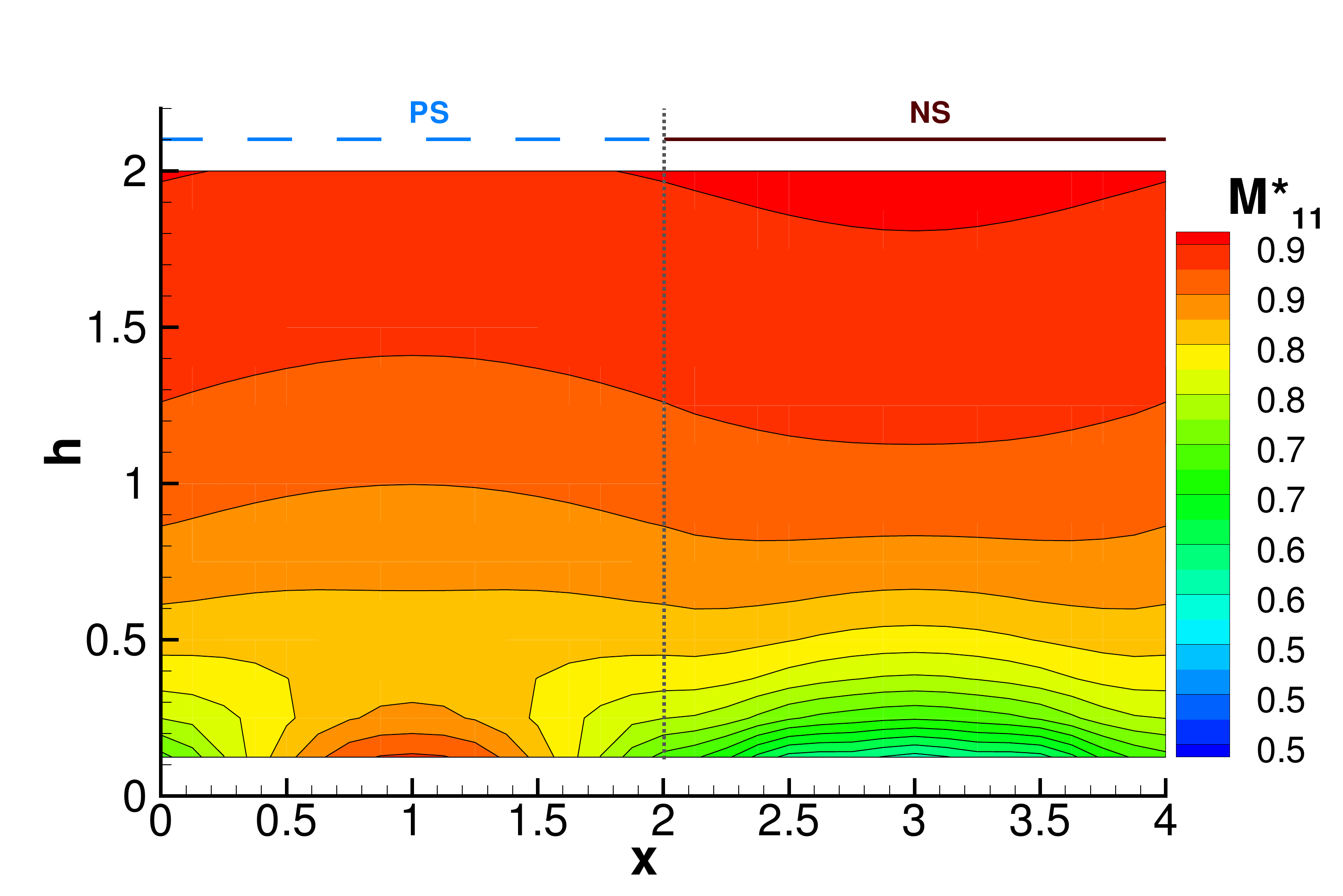}
\label{subfig:M11_0.25}
}
\subfigure[a=0.5]{
\includegraphics[width=0.5\textwidth]{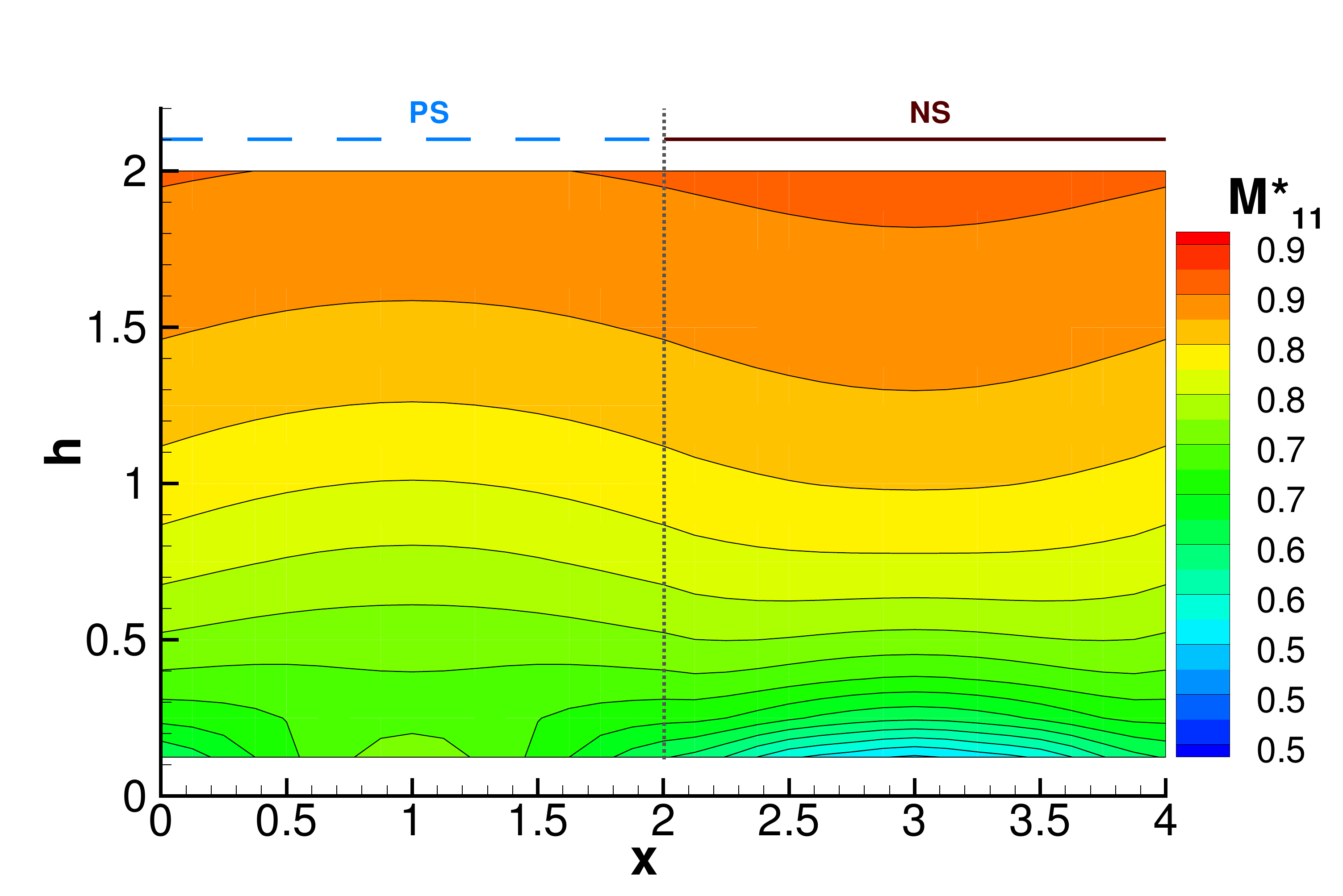}
\label{subfig:M11_0.5}
}
\subfigure[a=1]{
\includegraphics[width=0.5\textwidth]{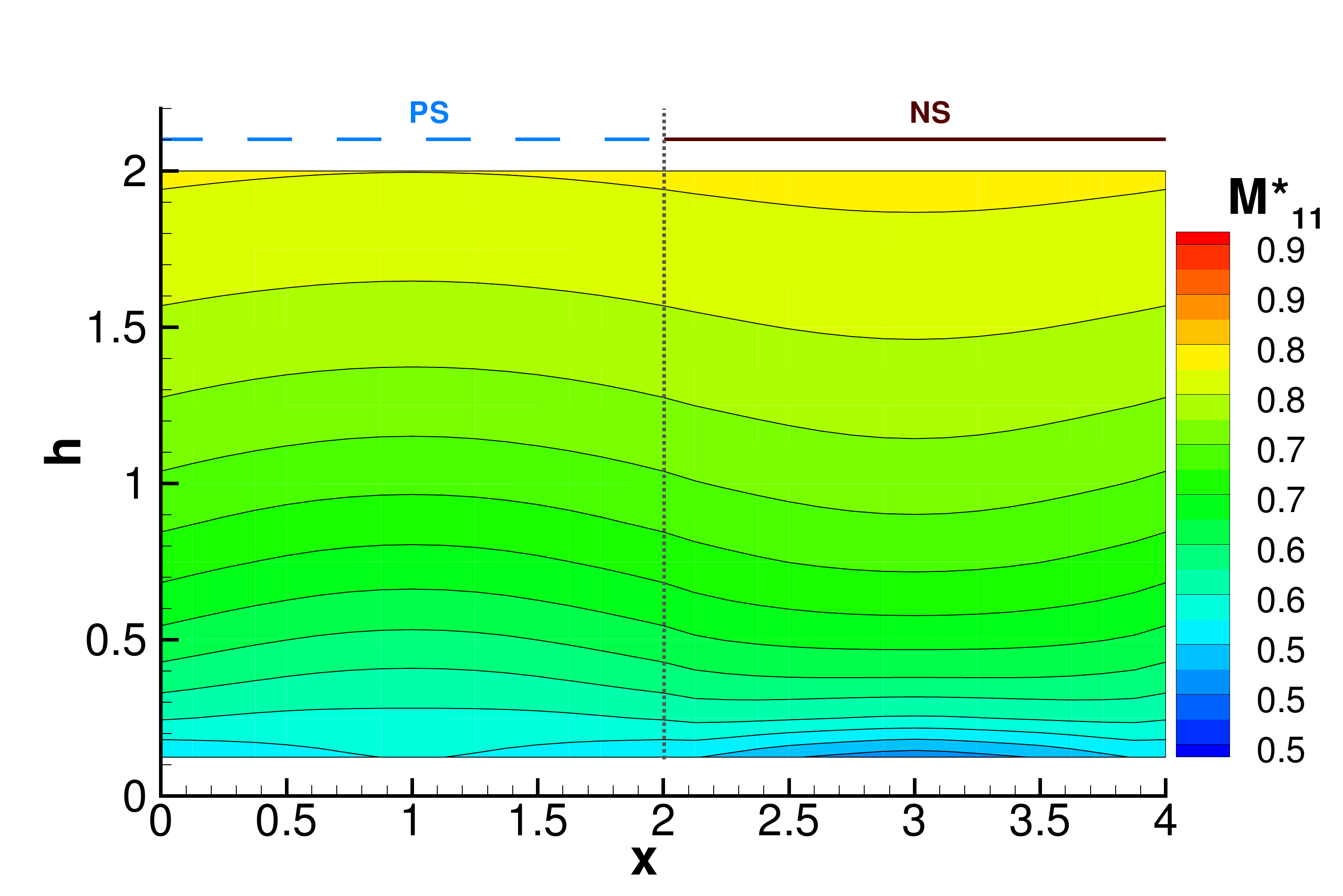}
\label{subfig:M11_1}
}
\subfigure[a=2]{
\includegraphics[width=0.5\textwidth]{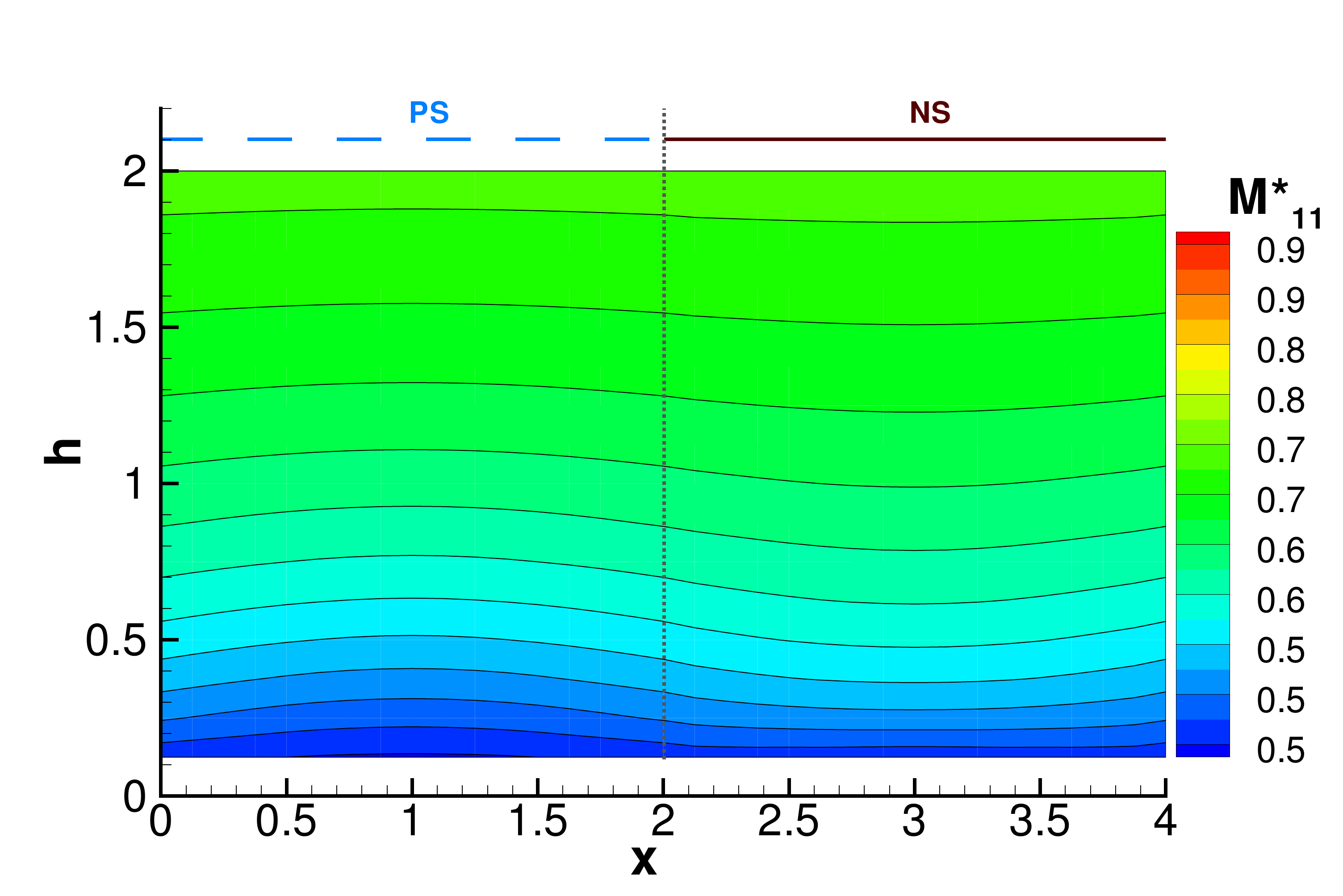}
\label{subfig:M11_2}
}
\caption{
$M^{*}_{11}$ fields for different sphere radii. 
Near the wall the mobility strongly
depends on the $x-$ position of the sphere center. The $x-$dependence is 
weakened as the gap $h$ is increased.
\label{fig:campi}
}
\end{figure*}
%%%%%%%%%%%%%%%%%%%%%%%%%%%%%%%%%%%%%%%%%%%%%%%%%%%%%%%%%%%%%%%%%%%%%%%%%%%%

\section{Mobility tensor for a superhydrophobic wall} \label{sec:mob_SH}

By symmetry, the matrix representing the mobility tensor of 
a sphere close to the superhydrophobic surface sketched in 
Fig.\ref{fig:system}, takes on a checkerboard structure. The (symmetric) 
matrix can easily  be recast into block-diagonal form by a simple reordering 
of generalized velocity and forces, namely
%%%%%%%%%%%%%%%%%%%%%%%%%%%%%%%%%%%%%%%%%%%%%%%%%%%%%%%%%%%%%%%%%%%%%%%%%%%%
\begin{equation}
\label{eq:Mblock}
\left[
\begin{array}{c}
U_1 \\
\omega_2 \\
U_3 \\
\omega_1 \\
U_2 \\
\omega_3 \\
\end{array}
\right] =
\left[
\begin{array}{cccccc}
M_{11} & M_{15} & M_{13} & 0 & 0 & 0\\
M_{15} & M_{55} & M_{35} & 0 & 0 & 0\\
M_{13} & M_{35} & M_{33} & 0 & 0 & 0\\
0 & 0 & 0 & M_{44} & M_{24} & M_{46} \\
0 & 0 & 0 & M_{24} & M_{22} & M_{26} \\
0 & 0 & 0 & M_{46} & M_{26} & M_{66} \\
\end{array}
\right]
\left[
\begin{array}{c}
F_1 \\
M_2 \\
F_3 \\
M_1 \\
F_2 \\
M_3 \\
\end{array}
\right].
\end{equation}
%%%%%%%%%%%%%%%%%%%%%%%%%%%%%%%%%%%%%%%%%%%%%%%%%%%%%%%%%%%%%%%%%%%%%%%%%%%%
In this form the cross-coupling between the different degrees of freedom becomes 
apparent showing, e.g., that the rotation around the axis parallel to the
stripes ($x_2$) couples with a force in the wall-normal direction $x_3$ through the mobility coefficient $M_{53}=M_{35}$. 
The mobility 
tensor depends on the wall-normal distance expressed by 
the gap clearance $h$ and on the wall-parallel coordinate  normal to the stripes $x_1=x$, $M_{\alpha \beta}(x,h)$.

In the following the (dimensionless)  mobility tensor is normalized by the free 
stream value $M_{11}^\infty = 1/(6 \pi a)$, $\vec{M}^*= \vec{M}/M_{11}^\infty$.
Figure~\ref{fig:campi} reports $M^*_{11}$ as a function of $h$ and $x$ for 
different particle radii $a=0.25,0.5,1,2$. Apparently $M^*_{11}$ is symmetric with 
respect to the center of both the perfect-slip ($x = 1$)  and the no slip ($x = 3$) 
stripe. Two features of the plots are noteworthy.  $i)$ Far from the wall
$M^*_{11}$ is unexpectedly larger for $x \in (2,4)$ (i.e. when the center of 
the sphere is above the no-slip stripe) than for $x \in (0,2)$ (sphere above the 
perfect-slip stripe). $ii)$ This behavior is reversed close to the wall where, as 
expected, $M^*_{11}$ is larger above the perfect slip stripe, see cases  
$a=0.25$ and $a=0.5$ in particular. 

\begin{figure}[t!]
\includegraphics[width=0.9\textwidth]{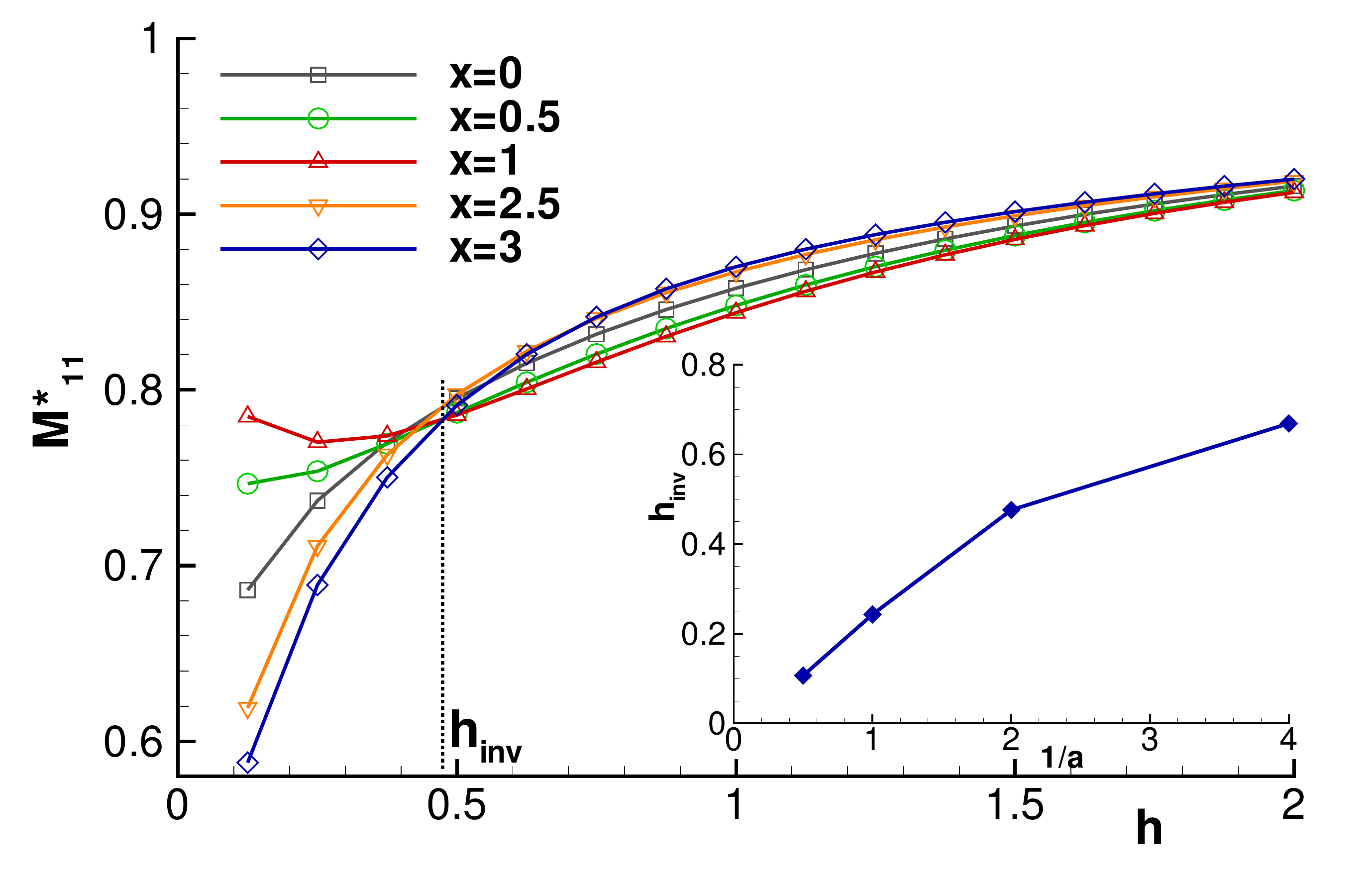}
\caption{$M^*_{11}$ as a function of the gap $h$ for different 
$x$ ($a=0.5$). Different behaviors are observed  in the near wall region 
$h < h_{inv}$ and in the far field $h > h_{inv}$. Near the wall strong 
changes occur when moving the particle from a perfect-slip stripe (e.g. $x=1$) 
to a  no-slip stripe (e.g. $x=3$) and the mobility is larger
above a perfect-slip  than above a no-slip stripe. 
In the far field  $M^*_{11}$  is more uniform in $x$ and the mobility
of the particle is larger just above a no-slip stripe. In the inset,
inversion gap $h_{inv}$ for different particle radii.
\label{fig:inversione}
}
\end{figure}

In Fig.~\ref{fig:inversione} $M^*_{11}$ is 
reported as a function of $h$ for different particle positions $x$ for $a=0.5$. 
Two distinct regions can be identified: a near-wall and a far field region. 
In the near wall region the mobility coefficient is larger for $x$ corresponding 
to the perfect slip stripe and strongly depends on the position along the 
wall pattern. In the far field the mobility is larger for $x$ corresponding to the 
no slip portion of the wall, with a less pronounced  $x$-dependence 
and a monotonic approach to the free space  value (recovered up to 90\% at $h=2$).  
In order to define the two regions, their boundary is set
at the gap $h_{inv}$  where $M^*_{11}(1,h_{inv}) =  M^*_{11}(3,h_{inv})$ 
(i.e. the gap where the mobility for a particle above the PS stripe 
equals the mobility  above the NS stripe). From the inset of 
Fig.~\ref{fig:inversione} it is apparent that $h_{inv}$ decreases with the particle 
radius. 

\begin{figure}[b!]
\includegraphics[width=0.9\textwidth]{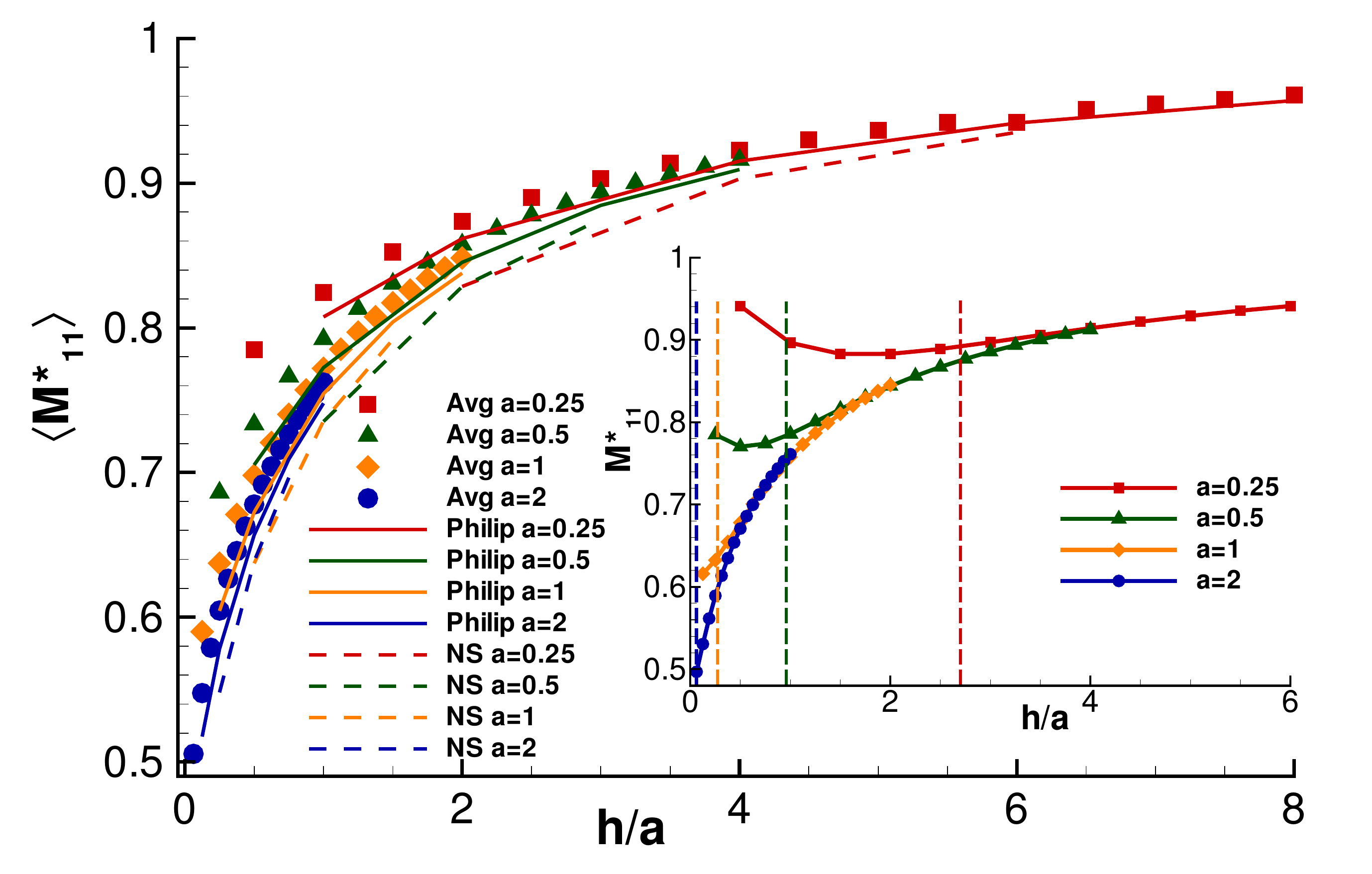}
\caption{Mobility profiles as a function of $h/a$ for particles of different 
radii $a\in[0.25;2]$. Spatially averaged mobility $\langle M^*_{11} \rangle$
for the patterned wall (symbols). Mobility of an homogeneous wall with
a partial slip boundary condition (solid lines). Non-slip 
homogeneous wall (dashed lines). Inset: $M^*_{11}$ profiles at the center of PS 
zone ($x = 1$). The dashed vertical lines denote the critical distance
$h_{inv}$.
}
\label{fig:invers_norm}
\end{figure}

Figure \ref{fig:invers_norm} shows the dimensionless averaged 
mobility coefficient $\langle M^*_{11} \rangle$ as a function of the 
gap-to-particle-radius ratio, $h/a$, for spheres of different radii. The 
angular brackets denotes the spatial average of the complete field
$M^*_{11}(x,h)$ in the periodic direction $x$. In the plot the present data (symbols) 
are compared against the mobility over a homogeneous wall with a suitably defined 
partial slip boundary condition (solid lines),
%%%%%%%%%%%%%%%%%%%%
\begin{equation}
\label{eqn:Navier_BC}
{u_{x/y}}|_w =  \ell_{x/y}  \frac{\partial u_{x/y}} {\partial z}\Big|_w  \ .
\end{equation}
%%%%%%%%%%%%%%%%%%%%
Philip formula \cite{philip1972flows,ng2010apparent} expresses the effective slip
lengths, $\ell_x$ and $\ell_y$, in the longitudinal and transversal directions to the stripes,
in terms of the pattern solid fraction  $\phi_s$ and  length $\lambda$,
%%%%%%%%%%%%%%%%%%%%
\begin{equation}
\label{eqn:slip_lenghts}
\ell_{y} = 2 \ell_x =  \frac{1}{\pi} \ln\left\{\sec\left[\frac{\pi \left(1-\phi_s \right)}{2} \right] \right\} \ .
\end{equation}
%%%%%%%%%%%%%%%%%%%%
In the original papers these expressions were derived for a flow over a patterned wall 
dragged by a constant shear stress in the far field.  The same expressions are used 
here to reconstruct an effective wall boundary condition to model the fluid-wall 
interaction in the case of the moving sphere over the patterned surface. In other 
words eqs.~(\ref{eqn:Navier_BC},\ref{eqn:slip_lenghts}) are used as boundary 
conditions at the patterned wall  in the BEM solver. Figure~\ref{fig:invers_norm} 
shows that, as $h/a$ is increased, the mobility approaches the free-space value 
independently of the detailed boundary condition used at the patterned wall. 
For further comparison also the data pertaining to a homogeneous no-slip wall are 
reported (dashed lines). The mobility profiles calculated with the partial-slip 
boundary condition closely follows those computed for the actually patterned wall 
for nearly all the considered gaps. Clearly the accuracy in reconstructing the correct 
mobility is better in the far field region, even though the discrepancy is always 
below  5\% in the worst cases when the sphere gets closest to the wall. 
These results indicate that the spatially average mobility experienced by the 
particle is weakly dependent on the geometrical details of the wall pattern and  can 
be described by a suitably defined  effective slippage at  the wall. 
The mobility profiles $M^*_{11}(x=1,h)$ extracted at the center of the  perfect-slip 
region are plotted  in the inset of Fig.~\ref{fig:invers_norm}.  
%%%%%%%%%%%%%%%%%%%%%%%%%%%%%%%%%%%%%%%%%%%%%%%%%%%%%%%%%%%%%%%%%%%%%%%%%%%%
%%%%%%%%%%%%%%%%%%%%%%%%%%%%%%%%%%%%%%%%%%%%%%%%%%%%%%%%%%%%%%%%%%%%%%%%%%%%
In the near wall region ($h < h_{inv}$) the mobility is strongly affected by the 
perfect-slip stripe and, at least for the spheres of radius 
$a=0.25,0.5$ (i.e. for a sphere diameter smaller than one stripe width), a mobility 
minimum is achieved at a certain (small) distance from the wall. The location of 
the minimum approaches the wall as the sphere radius is increased. In the far field 
($h > h_{inv}$, dashed vertical lines), the mobility approaches the free-space 
value closely following the curves obtained from the effective slip model of the wall.

%%%%%%%%%%%%%%%%%%%%%%%%%%%%%%%%%%%%%%%%%%%%%%%%%%%%%%%%%%%%%%%%%%%%%%%%%%%%
%%%%%%%%%%%%%%%%%%%%%%%%%%%%%%%%%%%%%%%%%%%%%%%%%%%%%%%%%%%%%%%%%%%%%%%%%%%%

\begin{figure}[t!]
\subfigure[$M^*_{22}$ field (a=0.5)]{
\includegraphics[width=0.5\textwidth]{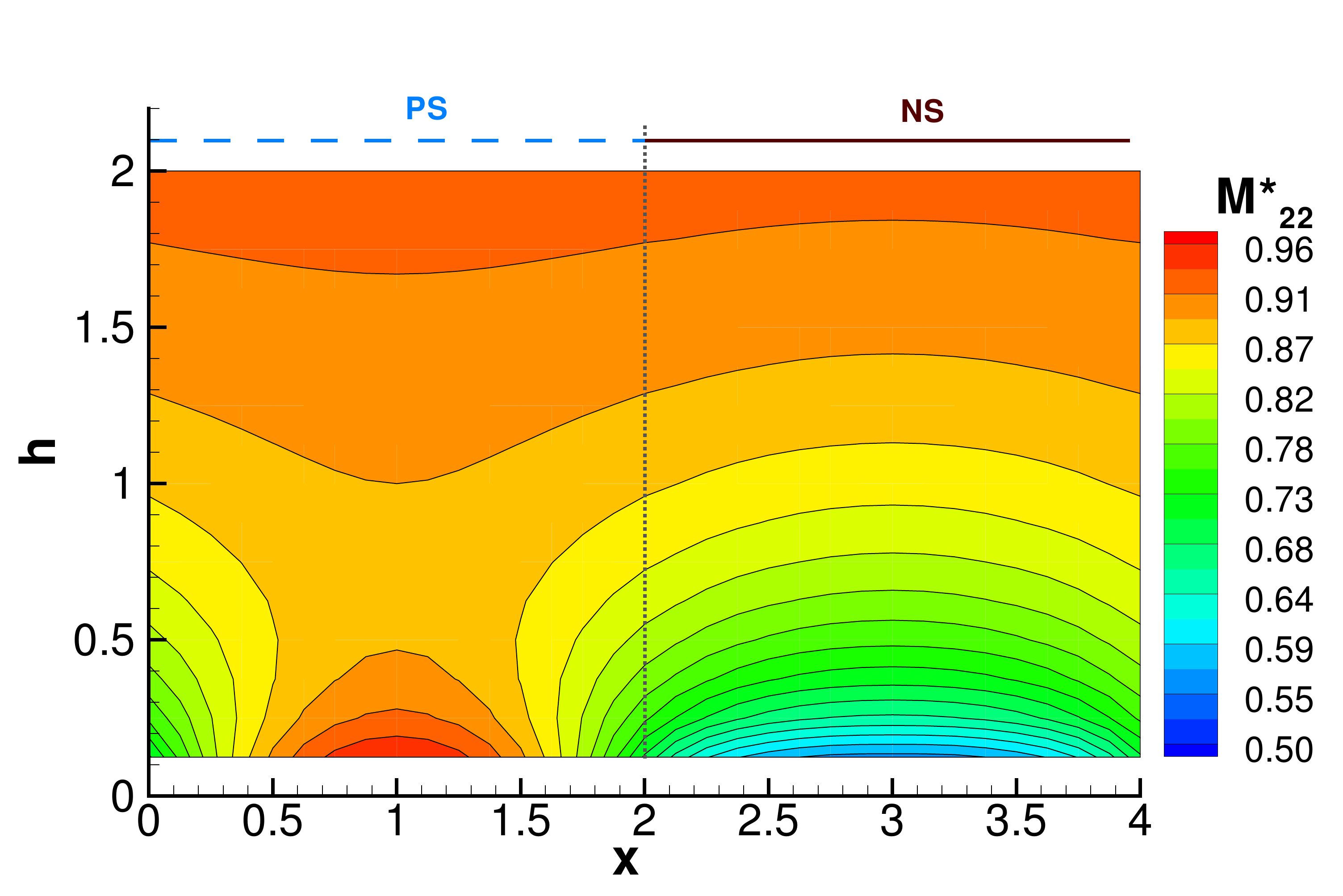}
\label{subfig:M22_r0.5}
}
\subfigure[$M^*_{13}$ field (a=0.5)]{
\includegraphics[width=0.5\textwidth]{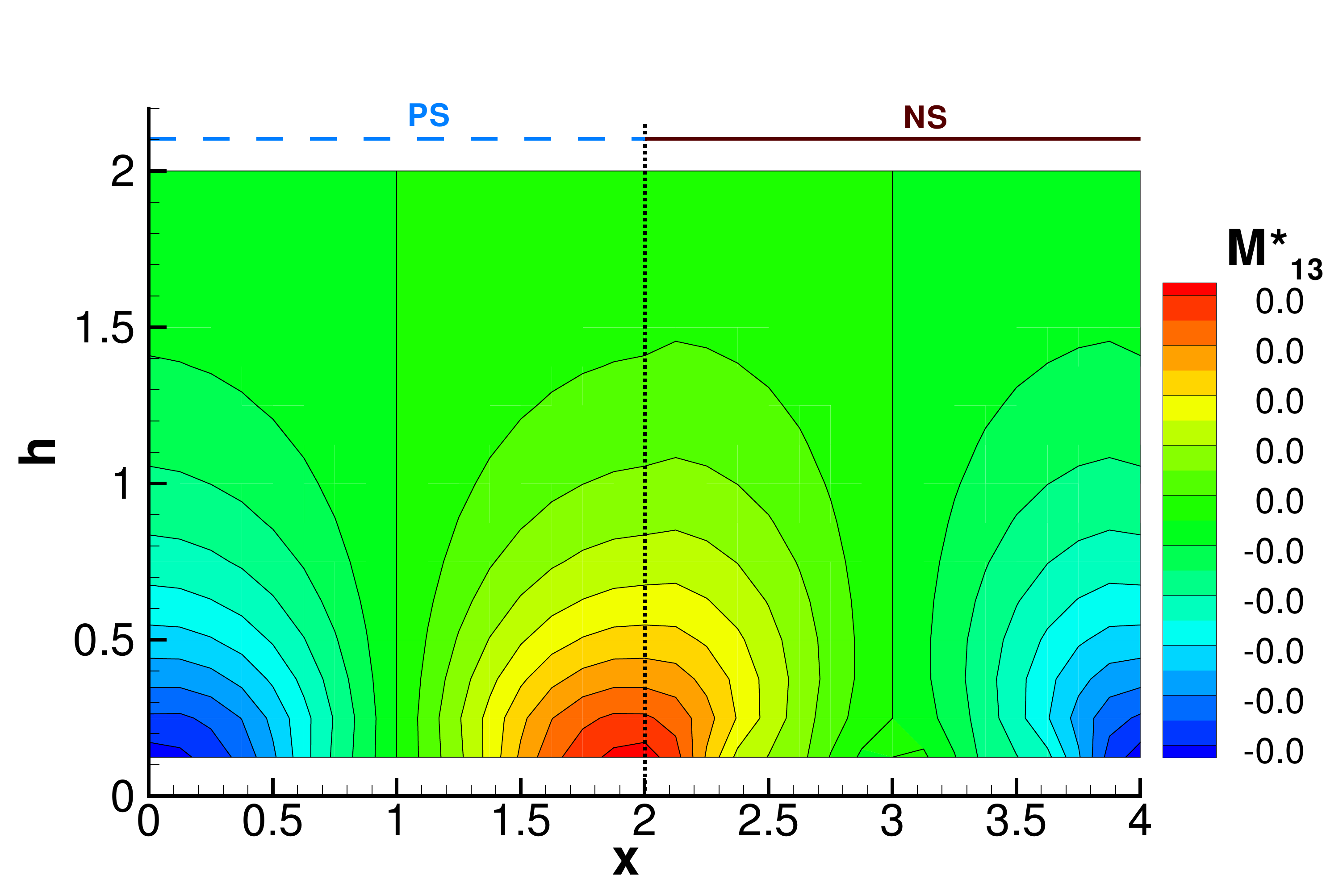}
\label{subfig:M13_r0.5}
}
\caption{$M^*_{22}$ (a) and  $M^*_{13}$ (b) vs $(x,h)$.
The maximum of $M^*_{22}$ above the PS region ($x=1$) and the minimum above 
NS ($x=3$) are apparent. $M^*_{13}$ is antisymmetric  with respect to the 
center of both the PS ($x=1$)  and the NS ($x=3$) region.
\label{fig:campi1}
}
\end{figure}

The discussion of the mobility is completed in Fig.~\ref{fig:campi1} which
provides coefficients $M^*_{22}$ and  $M^*_{13}$ for $a=0.5$.
As expected $M^*_{13}$ is anti-symmetric with respect to the mid-point of the 
two stripes. For comparison, in free space the mobility matrix is purely diagonal and isotropic 
meaning that there is no coupling among the degrees of freedom, 
$M_{11}=M_{22}=M_{33}$, and $M_{44}= M_{55}= M_{66}$.
A homogeneous wall in the $x-y$ plane breaks the homogeneity in the
$z$-direction, i.e. $\vec{M}=\vec{M}(h)$ and introduces the coupling between translations in 
wall parallel directions and rotations along the orthogonal wall-parallel 
axis, see the discussion at the end of \S~\ref{sec:num}. The coupling is provided by the  mobility coefficients 
$M_{15}(h) = M_{24}(h) = M_{51}(h) = M_{42}(h)$. For the homogeneous wall
the non vanishing terms are $M_{33}(h)$, $M_{11}(h)=M_{22}(h)$, 
$M_{15}(h) = M_{24}(h) = M_{51}(h) = M_{42}(h)$, $M_{66}(h)$, and 
$M_{44}(h)=M_{55}(h)$. 
%%%%%%%%%%%%%%%%%%%%%%%%%%%%%%%%%%%%%%%%%%%%%%%%%%%%%%%%%%%%%%%%%%%%%%%%%%%%
%%%%%%%%%%%%%%%%%%%%%%%%%%%%%%%%%%%%%%%%%%%%%%%%%%%%%%%%%%%%%%%%%%%%%%%%%%%%
Indeed, the non-vanishing coefficient  $M_{13}(x,h)$  shown in 
Fig.\ref{subfig:M13_r0.5} is a 
feature induced by the breaking of the $x$-invariance introduced by the pattern. 
More generally, as already discussed in connection with eq.~(\ref{eq:Mblock}), in 
presence of the stripes the mobility matrix splits into two $3 \times 3$ blocks.
One describes the coupling among $x$-translations, 
$y$-rotations axis and $z$-translations. The other block couples  
$y$-translations, $x$, and $z$-rotations. The existence of the non-vanishing 
coefficient $M_{31}=M_{13}$ in the first block implies that a force parallel 
to the wall and normal to the stripes generates a wall normal velocity  such 
that the issuing motion is no more contained in a wall parallel plane.	

\section{Application to particle separation} \label{sec:sorting}

The mobility field evaluated in the previous section can be 
exploited to compute the sphere trajectory under 
the action of an external force $\vec{F}$  parallel to the wall. 
The equation of motion reads
%%%%%%%%%%%%%%%%%%%%%%%%%%%%%%%%%%%%%%%%%%%%%%%%%%%%%%%%%%%%%%%%%%%%%%%%%%%%
\begin{equation}
\vec{\dot{x}}=\vec{M_{UF}}(x,h) \cdot \vec{F} \, ,
\label{eq:mfv} 
\end{equation}
%%%%%%%%%%%%%%%%%%%%%%%%%%%%%%%%%%%%%%%%%%%%%%%%%%%%%%%%%%%%%%%%%%%%%%%%%%%%
where $\vec{x} = (x,y,h)$ defines the sphere position and $\vec{M_{UF}}(x,h)$ 
is the upper $3 \times 3$ block of the mobility tensor $\vec{M}$ defined in 
eqs.~(\ref{eq:Mcompl}) and (\ref{eq:M}) where the subscripts $\vec U$ and 
$\vec F$ denote the linear velocity - force coupling. Rotations and torques are 
not explicitly addressed since they are irrelevant to the trajectory of the 
sphere center. The purpose is investigating the potential of SH surfaces in 
combination with suitable forms of external forcing to achieve particle 
separation, i.e. focusing particles with different characteristics in different 
regions of the flow domain.
\textcolor{black}{Although calculated for a single particle,  the present results 
can be used also for dilute suspensions. The limit concentration above which 
the results loose validity can be estimated by considering that the hydrodynamic 
interactions between neighboring spheres vanish as $1/r$, 
with $r$ the distance between the their centers. 
The results in~\cite{batchelor1976brownian,jeffrey1984calculation}
indicate that the interaction terms are negligible 
for $r > r_d  \propto \hat{a}$, where $\hat{a}$ is the average 
radius  in the particle suspension and the proportionality constant is  
order of a few tens. It follows a rough estimate for the particles concentration 
threshold above which hydrodynamic interaction matter,  $c \propto 1/(r_d)^3$.
Hence the present results can be consistently used also for dilute suspensions 
with a concentration $c \le c_d \propto r_d^{-3}$. 
}

In the following subsections no wall normal force is applied, $F_3 = 0$, while the 
wall parallel force normal to the stripes is taken to be constant, $F_1 = 1$.  
Different cases are considered as concerning the transversal force component, 
namely $F_2 = const$ in \S~\ref{sec:cf}, $F_2 = F_2(x)$  in \S~\ref{sec:fy_sinx}, 
and $F_2= F_2(t)$ in \S~\ref{sec:fy_sint}.

%%%%%%%%%%%%%%%%%%%%%%%%%%%%%%%%%%%%%%%%%%%%%%%%%%%%%%%%%%%%%%%%%%%%%%%%%%%%
\begin{figure}
\includegraphics[width=0.45\textwidth]{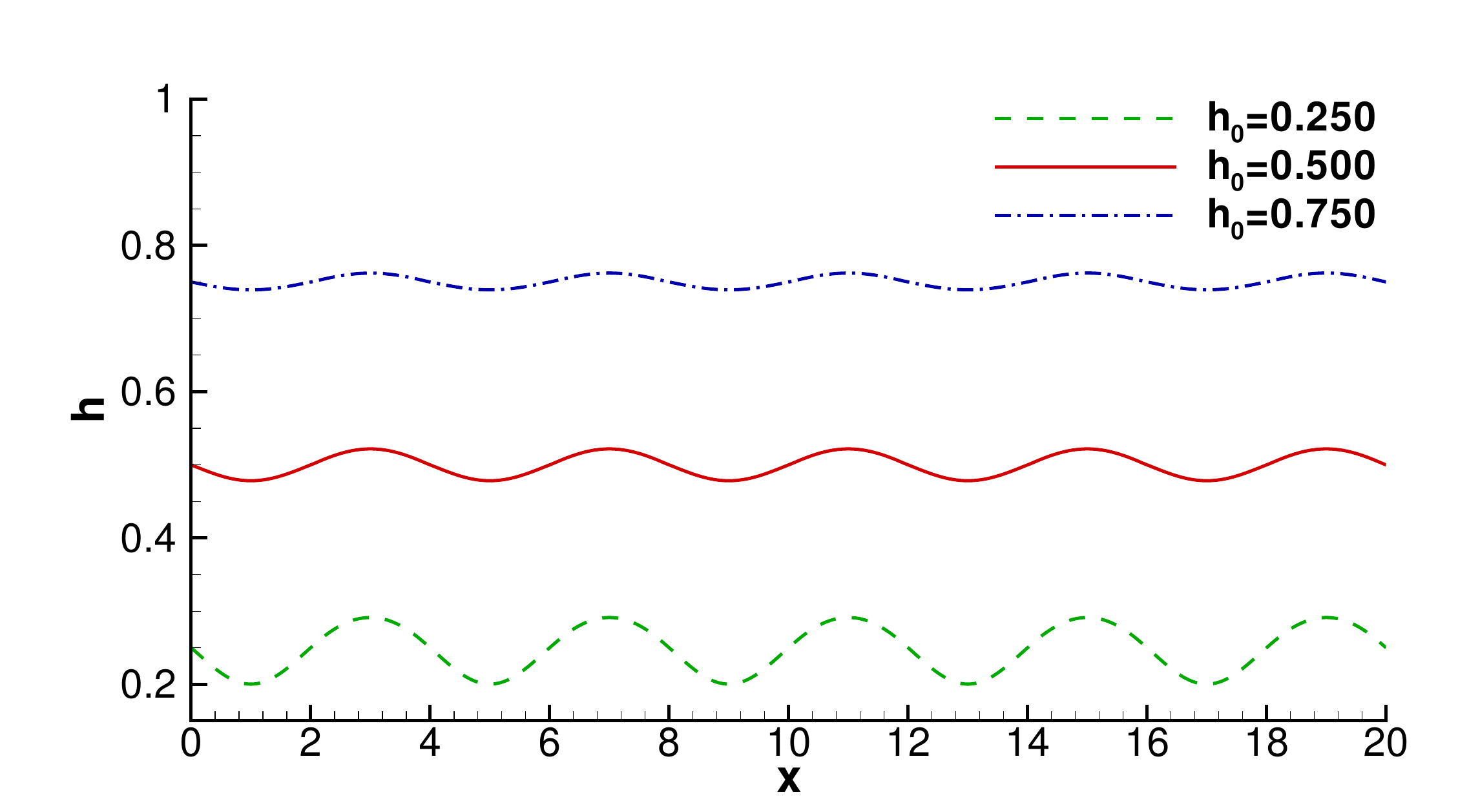}
\includegraphics[width=0.45\textwidth]{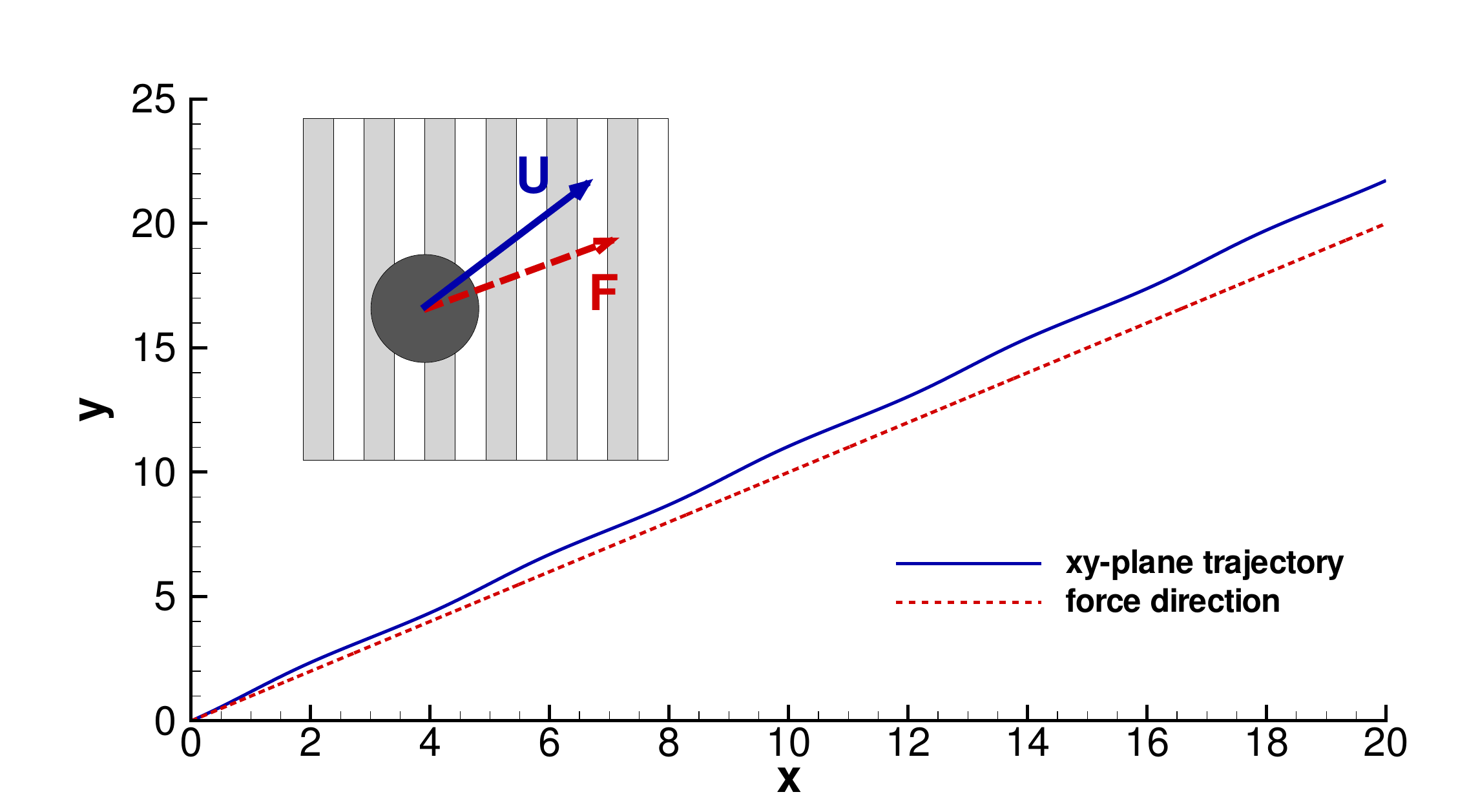} 
\caption{Trajectory in the $xh$ plane (top panel)
for a particle of radius $a=2$ released at $x_0 = 0$, $y_0 = 0$ 
for three different initial gaps: $h_0=0.25$ (dashed line), $h_0=0.5$ (solid line) and 
$h_0=0.75$ (dash-dotted line). The trajectory in the $xy$ for the case 
$h_0 = 0.25$ is reported in the bottom panel.
\label{fig:f_cost_tra}
}
\end{figure}
%%%%%%%%%%%%%%%%%%%%%%%%%%%%%%%%%%%%%%%%%%%%%%%%%%%%%%%%%%%%%%%%%%%%%%%%%%%%

%%%%%%%%%%%%%%%%%%%%%%%%%%%%%%%%%%%%%%%%%%%%%%%%%%%%%%%%%%%%%%%%%
\subsection{Constant forcing along the stripes} \label{sec:cf}

In the case of a constant force  with vanishing wall normal component, 
system (\ref{eq:mfv}) reduces to 
%%%%%%%%%%%%%%%%%%%%%%%%%%%%%%%%%%%%%%%%%%%%%%%%%%%%%%%%%%%%%%%%%%%%%%%%%%%%
\begin{eqnarray}
\frac{dx}{dt}&=&M_{11}(x,h) F_1 
\label{eq:f_cost1}\\
\frac{dy}{dt}&=&M_{22}(x,h) F_2 \
\label{eq:f_cost2}\\
\frac{dh}{dt}&=&M_{31}(x,h) F_1 \, .
\label{eq:f_cost3}
\end{eqnarray}
%%%%%%%%%%%%%%%%%%%%%%%%%%%%%%%%%%%%%%%%%%%%%%%%%%%%%%%%%%%%%%%%%%%%%%%%%%%%
In Fig.~\ref{fig:f_cost_tra} the $xh-$ and $xy-$projection of a typical 
trajectory are reported, top and bottom panel, respectively. The finite mobility 
coefficient $M_{31}$ couples a wall-normal motion to the wall-parallel force 
component $F_1$ acting along the stripe normal.
The wall normal motion occurs in periodic fashion with no mean drift. Typically, 
the oscillation amplitude is quite small, as expected given the small values 
of $M_{31}$. The motion in the wall-parallel plane $xy$ occurs at an average 
angle to the force direction, here oriented  $45^\circ$ from
the  stripe normal (x-direction). Both mean deflection 
and wall-normal oscillations are clearly induced by the alternating pattern of 
perfect- and no-slip stripes at the solid wall.

\begin{figure}
\includegraphics[width=0.45\textwidth]{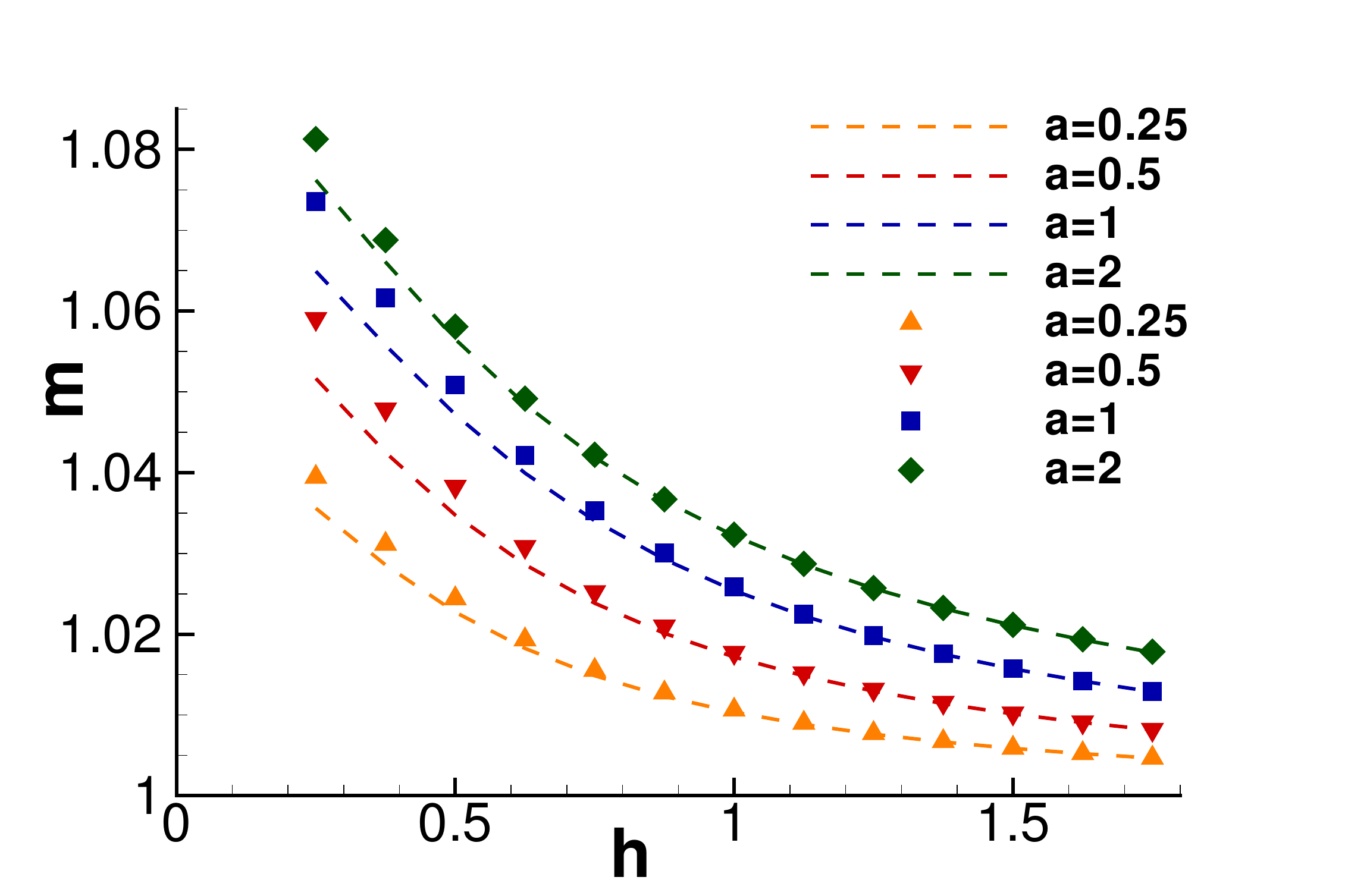}
\includegraphics[width=0.45\textwidth]{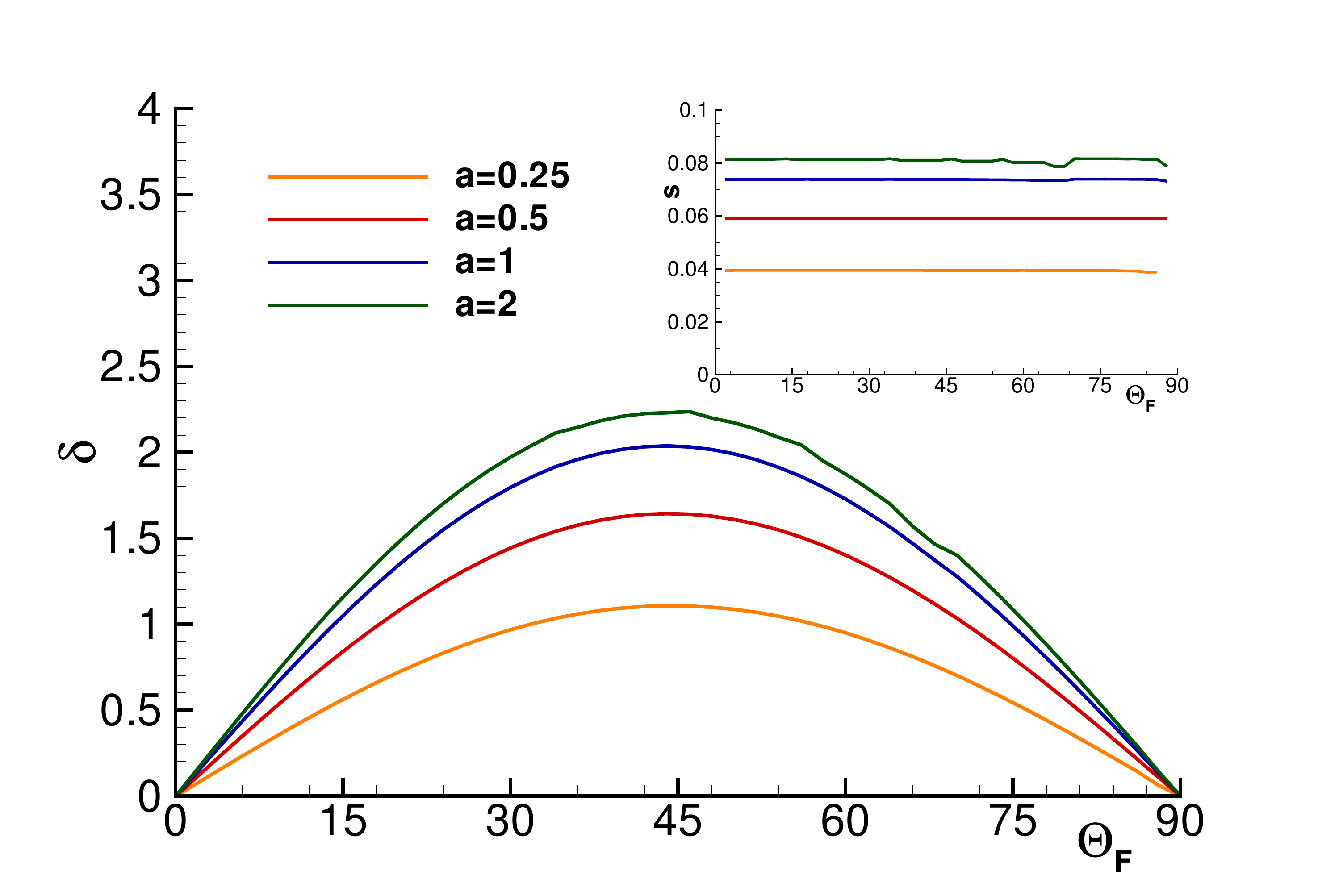}
\caption{Top panel:
Observed (symbols) and predicted (dashed lines) 
average slopes $m$  for particles 
driven by a $45^\circ$ constant forcing $F_2 / F_1 = 1$ as 
function of the initial gap $h_0$. 
\textcolor{black}{Bottom panel: Absolute deflection $\delta$ (degrees) with respect 
to the forcing angle $\Theta_F$  for $h_0 = 0.25$. The $45^\circ$ forcing achieves 
the maximum absolute deflection for a given radius. In the inset, measured 
relative difference between the actual trajectory slope $m$ and  tangent of 
the forcing angle slope $m_F = F_2/F_1$,  $s = (m-m_F)/m_F$.}
\label{fig:f_cost}
}
\end{figure}

The mean deflection observed in Fig.~\ref{fig:f_cost_tra} is worth being 
investigated in more detail, given its  potential interest for particle separation.
The deflection is measured by the average trajectory slope
$m = \langle dy /  dx \rangle$, where angular brackets denote averaging 
over the spatial period $\lambda$ of the stripes. 
The slope $m$ is reported in Fig.~\ref{fig:f_cost} (top panel)
for different particle radii $a$ and initial positions $(0,0,h_0)$ for a 
$45^\circ$ inclination of the force, $F_2/F_1 = 1$. 
The behavior illustrated in the figure is generic and does not depend on the force 
inclination nor on the particle initial position along the stripe pattern.
By inspection of the data the deflection increases as the gap $h$ is reduced 
and as the particle radius $a$ is augmented. 
\textcolor{black}{This not trivial behavior is induce by 
the elliptic nature of the Stokes equations. However,
its physical interpretation can be at least sketched for particles with radius much smaller the
the stripe width, $a \ll \lambda$. Far from the border 
between  perfect- and no-slip regions, the small
particle experiences a locally isotropic wall  
(i.e. $M_{11} = M_{22}$), implying no deflection. However, in a region of 
characteristic size $\epsilon \propto a$ straddling the border,  
$M_{11}$ and $M_{22}$ differ significantly inducing a local deflection of 
the particle path.
The overall deflection is somehow a weighted average between the non deflecting 
portions of the trajectory and the deflection regions
located at the stripe borders. By increasing the particle radius $a$,  
the size of the non deflecting portions decreases. At the same time the 
deflecting effect of the border gets stronger, since it scales only with the 
ratio $a/h$ under the assumption $a\ll\lambda$. It follows that the overall 
deflection increases with $a$.
This argument, even though strictly valid only in the limit 
$a \ll \lambda$,  can provide a guideline in interpreting the general behavior.
}

Clearly the amount of deflection does indeed depend on the
forcing directions and vanishes for forces perfectly normal or 
parallel to the stripes.  In any case, the maximum 
deflection  turns out to be small, at most a few percent with respect to the 
forcing angle. This finding is consistent with similar results discussed 
in~\cite{zhang2012separation} where molecular dynamics is used to compute the 
trajectory of a particle dragged by the underlying flow and 
in~\cite{zhou:194706} where a Poiseuille flow over a patterned wall 
is considered.

Given  the slope of the particle trajectory in the xy plane,
%%%%%%%%%%%%%%%%%%%%%%%%%%%%%%%%%%%%%%%%%%%%%%%%%%%%%%%%%%%%%%%%%%%%%%%%%%%%
\begin{equation}
\frac{dy}{dx}=\frac{M_{22}(x,h)}{M_{11}(x,h)} \frac{F_2}{F_1} \, ,
\label{eq:slope}
\end{equation}
%%%%%%%%%%%%%%%%%%%%%%%%%%%%%%%%%%%%%%%%%%%%%%%%%%%%%%%%%%%%%%%%%%%%%%%%%%%%
a rough approximation for the average  slope based on the small oscillations of 
$h(x)$ is
%%%%%%%%%%%%%%%%%%%%%%%%%%%%%%%%%%%%%%%%%%%%%%%%%%%%%%%%%%%%%%%%%%%%%%%%%%%%
\begin{equation}
m  
= 
\left \langle \frac{dy}{dx} \right \rangle
\simeq
\left \langle \frac{M_{22}(x,h_0)}{M_{11}(x,h_0)} \right \rangle
\frac{F_2}{F_1} \, .
\label{eq:slopec}
\end{equation}
%%%%%%%%%%%%%%%%%%%%%%%%%%%%%%%%%%%%%%%%%%%%%%%%%%%%%%%%%%%%%%%%%%%%%%%%%%%%
The above  estimate corresponds quite well to the actual data, as shown 
in Fig.~\ref{fig:f_cost} (top panel) by the comparison of symbols and dashed lines.
Significant discrepancies become apparent  for small gaps $h_0$, where 
$M_{31}(x,h)$ becomes significant (see Fig.~\ref{fig:campi1}) and the 
wall-normal oscillation is no more negligible, see the top panel of 
Fig.~\ref{fig:f_cost_tra}.

%%%%%%%%%%%%%%%%%%%%%%%%%%%%%%%%%%%%%%%%%%%%%%%%%%%%%%%%%%%%%%%%%%%%%%%%%%%%
%%%%%%%%%%%%%%%%%%%%%%%%%%%%%%%%%%%%%%%%%%%%%%%%%%%%%%%%%%%%%%%%%%%%%%%%%%%%
\textcolor{black}{
The bottom panel of Fig.~\ref{fig:f_cost} 
shows the absolute angular deflection $\delta = \Theta - \Theta_F$ 
with $\Theta = \tan^{-1}(m)$ and $\Theta_F = \tan^{-1}(m_F)$, where 
$m_F = F_2/F_1$.
A maximum angular deflection is apparent for  $\Theta_F=45^\circ$
while, as expected by symmetry consideration, $\delta = 0 $ for
$\Theta_F = 0^\circ$ (perpendicular to the stripes) 
and $\Theta_F = 90^\circ$  (parallel).
Interestingly, defined the  relative slope difference  
%%%%%%%%%%%%%%%%%%%%%%%%%%%%%%%%%%%%%%%%%%%%%%%%%%%%%%%%%%%%%%%%%%%%%%%%%%%%
\begin{equation}
s  
=
\frac{m - m_F}{m_F}  \, ,
\label{eq:rel_slope}
\end{equation}
%%%%%%%%%%%%%%%%%%%%%%%%%%%%%%%%%%%%%%%%%%%%%%%%%%%%%%%%%%%%%%%%%%%%%%%%%%%%
eq. (\ref{eq:slopec}) implies that 
$s \simeq \langle M_{22}(x,h_0) /M_{22}(x,h_0 \rangle - 1$,
i.e. $s$ does not depend on the direction of the applied 
force. 
This is consistent with the data reported in the inset of the bottom panel 
of Fig.~\ref{fig:f_cost}. 
}

\subsection{Spatially periodic forcing along the stripes} 
\label{sec:fy_sinx}

When the transverse force component $F_2$ is an oscillating function of the 
coordinate $x$ normal to the stripes, certain resonance effects may emerge. 
To address the problem, one may exploit system (\ref{eq:f_cost1}-\ref{eq:f_cost3})
by noticing that the equations for  $x(t)$ and $h(t)$ are $y$-independent. 
Hence the system can be first solved for the unknowns $x(t)$ and $h(t)$ for a 
constant force $F_1=1$, $F_3=0$, as in \S~\ref{sec:cf}. 
The solution of eq.~(\ref{eq:f_cost2}) then provides $y(t)=y[t;x,h,F_2]$ as a 
functional of $x(t), h(t)$ and $F_2(x)$. This functional  is linear in $F_2(x)$.
The average deflection of the particle along the stripes per wavelength of wall 
pattern will then depend on amplitude and shape of $F_2(x)$. The force distribution can be optimized
to maximize the  particle drift for given physical constraints, e.g. periodic, zero average force $\langle F_2\rangle = 0$ and prescribed  effective amplitude $A =\sqrt{\langle F_2^2 \rangle}$. It is not difficult to show 
that the solution of the optimization problem is
%%%%%%%%%%%%%%%%%%%%%%%%%%%%%%%%%%%%%%%%%%%%%%%%%%%%%%%%%%%%%%%%%%%%%%%%%%%%
\begin{equation}
\label{eq:opt} 
F_2^{opt}(x) = 
A 
\frac 
{\frac{\displaystyle M_{22}\left[x,h(x)\right]}{\displaystyle M_{11}\left[x,h(x)\right]F_1}- 
\frac{\displaystyle 1}{\displaystyle \lambda}\left\langle\frac{\displaystyle  M_{22}}{\displaystyle  (M_{11}F_1)}\right\rangle}
{\sqrt
  {\left \langle 
  \left( \frac{\displaystyle M_{22}}{\displaystyle M_{11}F_1}\right)^2
  \right \rangle - 
  \frac{\displaystyle 1}{\displaystyle \lambda}\left\langle \frac{\displaystyle M_{22}}{\displaystyle M_{11}F_1}\right\rangle^2 }
}  \, , 
\end{equation}
%%%%%%%%%%%%%%%%%%%%%%%%%%%%%%%%%%%%%%%%%%%%%%%%%%%%%%%%%%%%%%%%%%%%%%%%%%%%
where $h(x) = h\left[t(x) \right]$ with $t(x)$ the inverse of $x(t)$. 
From eq.~(\ref{eq:slope}) the optimized slope follows as
%%%%%%%%%%%%%%%%%%%%%%%%%%%%%%%%%%%%%%%%%%%%%%%%%%%%%%%%%%%%%%%%%%%%%%%%%%%%
\begin{equation}
\label{eq:m_opt} 
m^{opt} = 
\frac{A}{\lambda F_1} 
\sqrt{ 
\left\langle \frac{M_{22}}{M_{11}F_1}^2 \right\rangle - 
\frac{1}{F_1 \lambda}
\left\langle \frac{ M_{22}}{M_{11}F_1} \right\rangle^2 
} \ .
\end{equation}
%%%%%%%%%%%%%%%%%%%%%%%%%%%%%%%%%%%%%%%%%%%%%%%%%%%%%%%%%%%%%%%%%%%%%%%%%%%%

\begin{figure}[t!]
\includegraphics[width=0.8\textwidth]{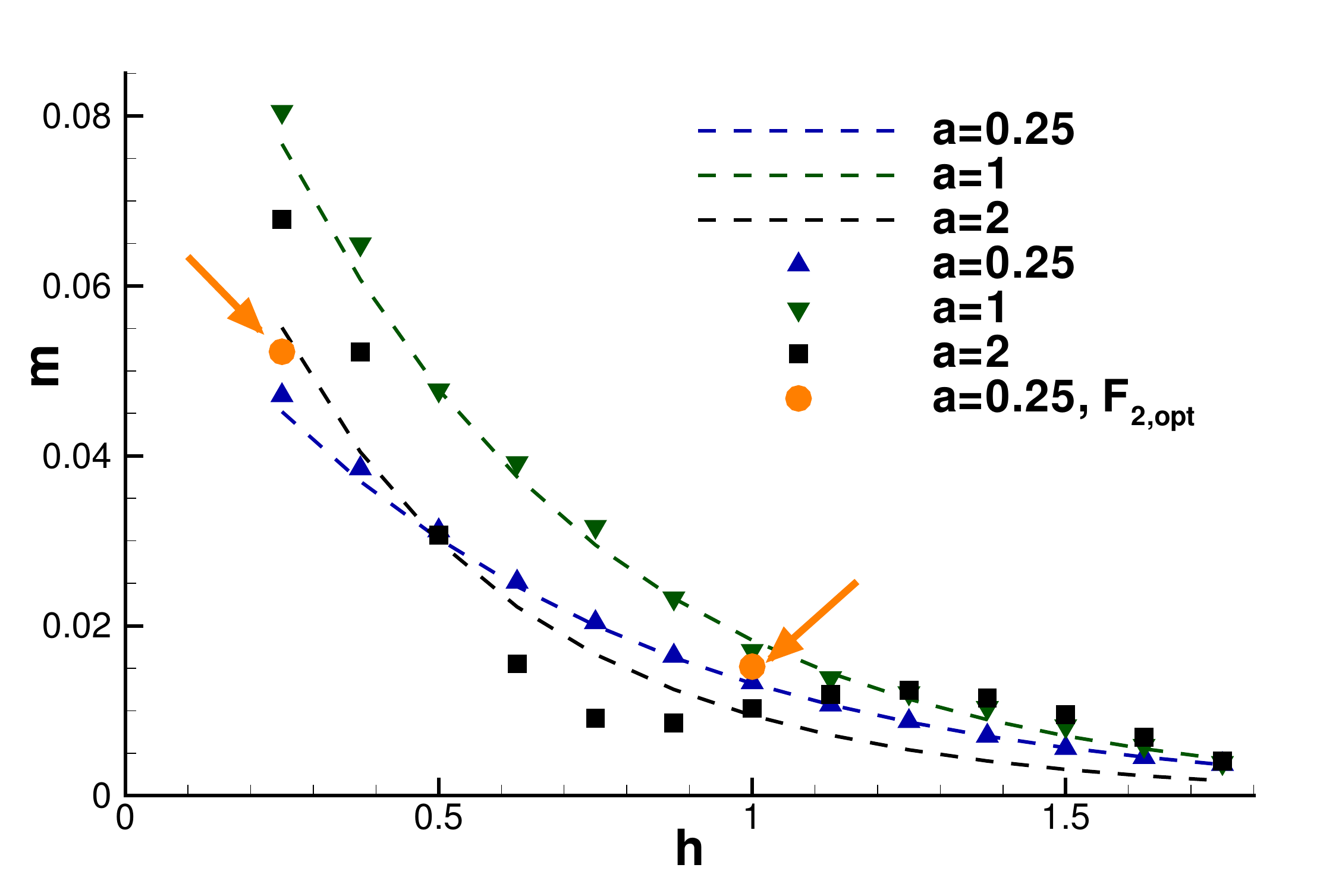}\\
\includegraphics[width=0.8\textwidth]{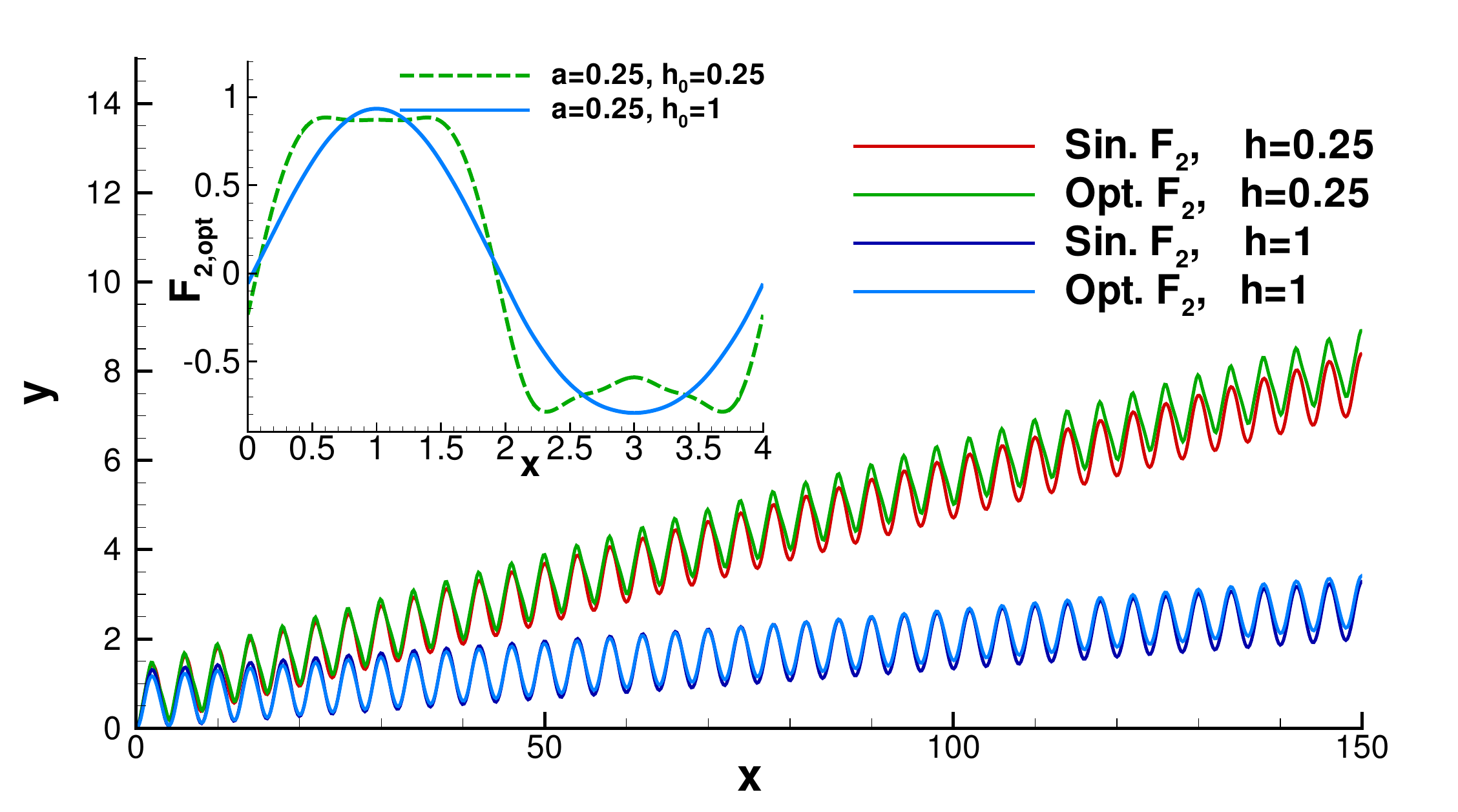}
\caption{
Top panel: observed (symbols) and predicted (dashed lines) average slopes $m$ for 
particles driven by a constant force in the $x$ direction ($F_1 = 1$) 
and $x-$dependent force in the $y-$direction (eq.~\ref{eq:fy_sin_x})
with $A=\sqrt{2}$ as function of the initial gap $h_0$. Circles are referred to 
average slopes $m$ obtained using the optimal forcing in the $y-$direction 
(eq.~\ref{eq:opt}). Bottom panel: comparison of trajectories obtained by 
sinusoidal and optimal forcing in the $y-$direction. 
\textcolor{black}{The top curve pair compares the trajectories 
at $h=0.25$. The bottom pair provides the comparison at $h=1$, where the 
two curves are almost superimposed. }
The inset shows $F_2^{opt}(x)$ (eq.~\ref{eq:opt}) for two values $h_0=0.25,1$.
\label{fig:f_yosci}
}
\end{figure}

Instead of using the solution of the  optimization problem, 
in the following $F_2(x)$  is taken to be a sinusoid,
%%%%%%%%%%%%%%%%%%%%%%%%%%%%%%%%%%%%%%%%%%%%%%%%%%%%%%%%%%%%%%%%%%%%%%%%%%%%
\begin{equation}
\label{eq:fy_sin_x} 
\frac{dy}{dt} =
M_{22}(x,h) \frac{A}{\sqrt{2}} \sin\left(\frac{2\pi x}{\lambda}\right) \, ,
\end{equation}
%%%%%%%%%%%%%%%%%%%%%%%%%%%%%%%%%%%%%%%%%%%%%%%%%%%%%%%%%%%%%%%%%%%%%%%%%%%%
a simple shape that is, in many cases, not too far from the optimal. 
Indeed, at least for trajectories not too close to the surface, 
the two mobility coefficients are reasonably well approximated by the expressions 
$M_{22/11} \simeq M_{22/11}^0+M_{22/11}^1 \sin\left(2 \pi x/\lambda \right)$, 
implying from eq.~(\ref{eq:opt}) 
$F_2^{opt} \sim  \sin\left(\frac{2\pi x}{\lambda}\right) $.
Figure~\ref{fig:f_yosci} (top panel) shows the measured values of $m$ as a function of the initial gap 
$h_0$ for several sphere radii. The slope $m$ is calculated as an  
average along an actual trajectory, $m=\langle dy/dx \rangle$. For relatively small
values of $a$ the behavior of $m$ vs $h_0$ is monotonic. As the sphere reaches the 
size $a =2$ (i.e. its diameter equals the pattern period) 
a new behavior emerges. This is presumably associated with
the strong oscillations in  $h(t)$ that occur in these conditions, related
to the increased $M_{13}$. 
$m$ scales linearly with the amplitude of the  forcing such 
that $m/A$ depends only on the initial gap $h_0$ and on the particle radius.

A rough estimate of the deflection coefficient $m$, at least for small oscillations in $h$, is
%%%%%%%%%%%%%%%%%%%%%%%%%%%%%%%%%%%%%%%%%%%%%%%%%%%%%%%%%%%%%%%%%%%%%%%%%%%%
\begin{equation}
m  = 
\left \langle \frac{dy}{dx} \right \rangle
\simeq
\left \langle \frac{M_{22}(x,h_0) \sin({2\pi x}/{\lambda})}
{M_{11}(x,h_0)} \right \rangle
\frac{A}{F_1} \, .
\label{eq:slopecsin}
\end{equation}
%%%%%%%%%%%%%%%%%%%%%%%%%%%%%%%%%%%%%%%%%%%%%%%%%%%%%%%%%%%%%%%%%%%%%%%%%%%%
The results of this approximation (dashed lines) compare  well with the data 
shown by symbols  in Fig.~\ref{fig:f_yosci} for  $a < 2$.  At $a=2$, the 
accuracy is lost due to the strong oscillations in $h$ which spoil the approximation.
%%%%%%%%%%%%%%%%%%%%%%%%%%%%%%%%%%%%%%%%%%%%%%%%%%%%%%%%%%%%%%%%%%%%%%%%%%%%
%%%%%%%%%%%%%%%%%%%%%%%%%%%%%%%%%%%%%%%%%%%%%%%%%%%%%%%%%%%%%%%%%%%%%%%%%%%%

As a last remark,  the trajectory under the optimal force $F_2^{opt}(x)$, eq.~(\ref{eq:opt}), is compared to
the results with the sinusoidal force,  bottom panel of figure \ref{fig:f_yosci} for $a=0.25$. 
Two cases are considered, namely $h_0=0.25$ and  $h_0=1$.  As expected, far from the
wall, i.e. at $h_0=1$, the two forces are almost equivalent. In the near wall region
the optimal force is indeed slightly more effective  in maximizing the drift. 
The shape of the optimal force is compared for the wall-distances
$h_0=0.25$ and $1$ in the inset of Fig.~\ref{fig:f_yosci}. At large distance the optimal force approaches a sinusoid,
while substantial differences are found closer to the wall where the optimal shape
resembles a square-wave consistently with the discontinuous boundary conditions enforced at surface. 
Even in this case the comparison of the trajectories confirms that the sinusoid is nearly optimal in maximizing the drift.
The  drift coefficients $m$ under optimal forcing are reported in the top panel
of figure \ref{fig:f_yosci} as filled circles highlighted by arrows.\\

\begin{figure}
\includegraphics[width=0.9\textwidth]{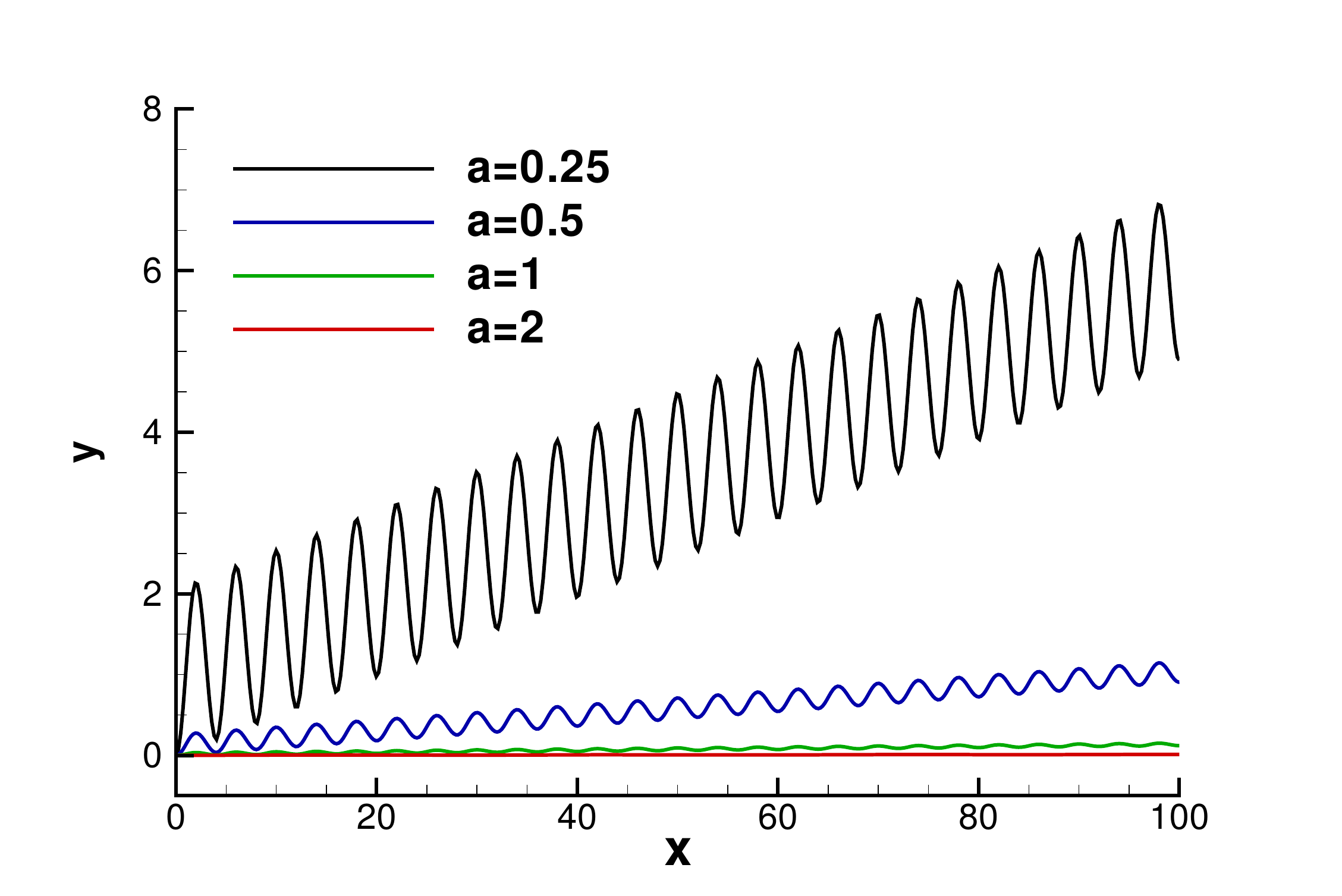}
\caption{
Trajectories obtained for a sinusoidal forcing (eq.~\ref{eq:fy_sin_x}) for 
different particle radii, when the external force $F_1$ in the stripe-normal 
direction is proportional to the particle volume like, e.g. in the case of the 
gravity force. The smaller the particle, the slower its stripe-normal motion 
\textcolor{black}{(curves from top to bottom correspond to particles with increasing radii):
the transversal force $F_2$ acts for a longer time on the smaller spheres  producing
larger oscillations amplitudes and mean deflection in the trajectory.} 
\label{fig:gravita}
}
\end{figure}

It might be interesting to specialize the traction force $F_1$ for the very 
common case where the force is proportional to the particle volume, 
$F_1 \propto \frac{4}{3}\pi a^3$, like it happens, e.g., for the vertical settling 
of a particle under gravity. The cubic dependence on the particle radius 
results in a strong reduction of the velocity normal to the horizontal stripes 
(crossing velocity) for smaller particles. The increased crossing time $T_c$ 
leads to a larger transversal impulse $\int_0^{T_c} F_2 dt$, thus amplifying the 
oscillations and the mean deflection of the particle. Indeed,
the differences between small and large particle radii apparent in 
Fig.~\ref{fig:gravita} may inspire simple particle sorting 
devices based on the coupling of the radius-dependent traction force with the transversal oscillating field.

%%%%%%%%%%%%%%%%%%%%%%%%%%%%%%%%%%%%%%%%%%%%%%%%%%%%%%%%%%%%%%%%%%%%%%%%%%%%

%%%%%%%%%%%%%%%%%%%%%%%%%%%%%%%%%%%%%%%%%%%%%%%%%%%%%%%%%%%%%%%%%%%%%%%%%%%%

\subsection{Time dependent forcing along the stripes}
\label{sec:fy_sint}

The previous section showed that significant mean particle drifts can be 
achieved by an oscillating spatial distribution of the transverse force. 
This finding suggests to look at the effect of a time-oscillating 
space-homogeneous transverse force $F_2(t)$. As in the previous section the 
particle is pulled along the $x$ direction by a constant force $F_1=1$ with 
a superimposed transverse oscillating forcing with zero mean. 
The resulting equation for the transverse motion of the particle is
%%%%%%%%%%%%%%%%%%%%%%%%%%%%%%%%%%%%%%%%%%%%%%%%%%%%%%%%%%%%%%%%%%%%%%%%%%%%
\begin{equation}
\frac{dy}{dt}=M_{22}(x,h) B \sin\left(\frac{2\pi t}{T}\right)  \, ,
\label{eq:oscit1}
\end{equation}
%%%%%%%%%%%%%%%%%%%%%%%%%%%%%%%%%%%%%%%%%%%%%%%%%%%%%%%%%%%%%%%%%%%%%%%%%%%%
to be solved together with equations (\ref{eq:f_cost1}) and (\ref{eq:f_cost3}).
%%%%%%%%%%%%%%%%%%%%%%%%%%%%%%%%%%%%%%%%%%%%%%%%%%%%%%%%%%%%%%%%%%%%%%%%%%%%
\begin{figure}[b!]
\includegraphics[width=0.8\textwidth]{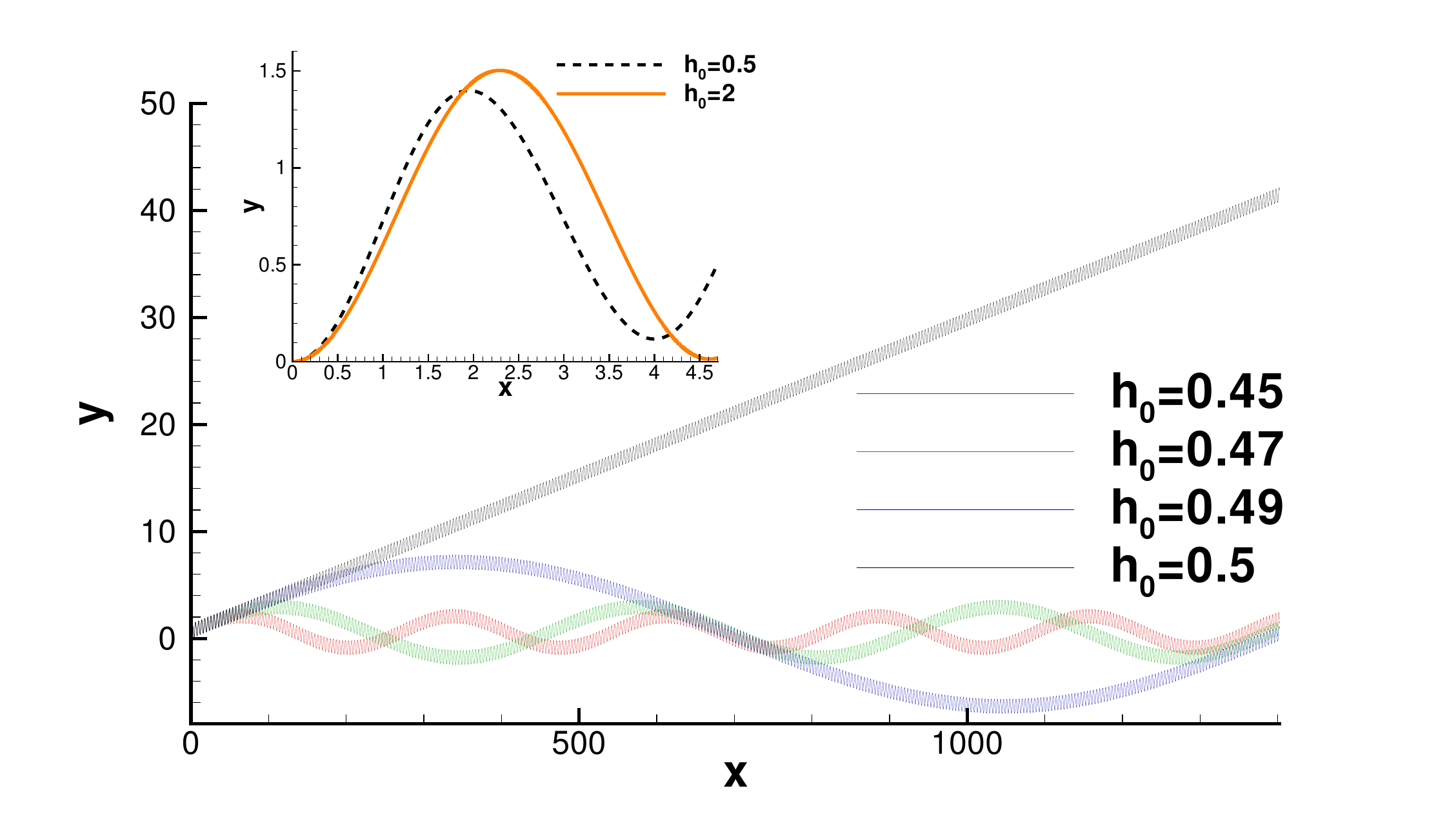}\\
\includegraphics[width=0.8\textwidth]{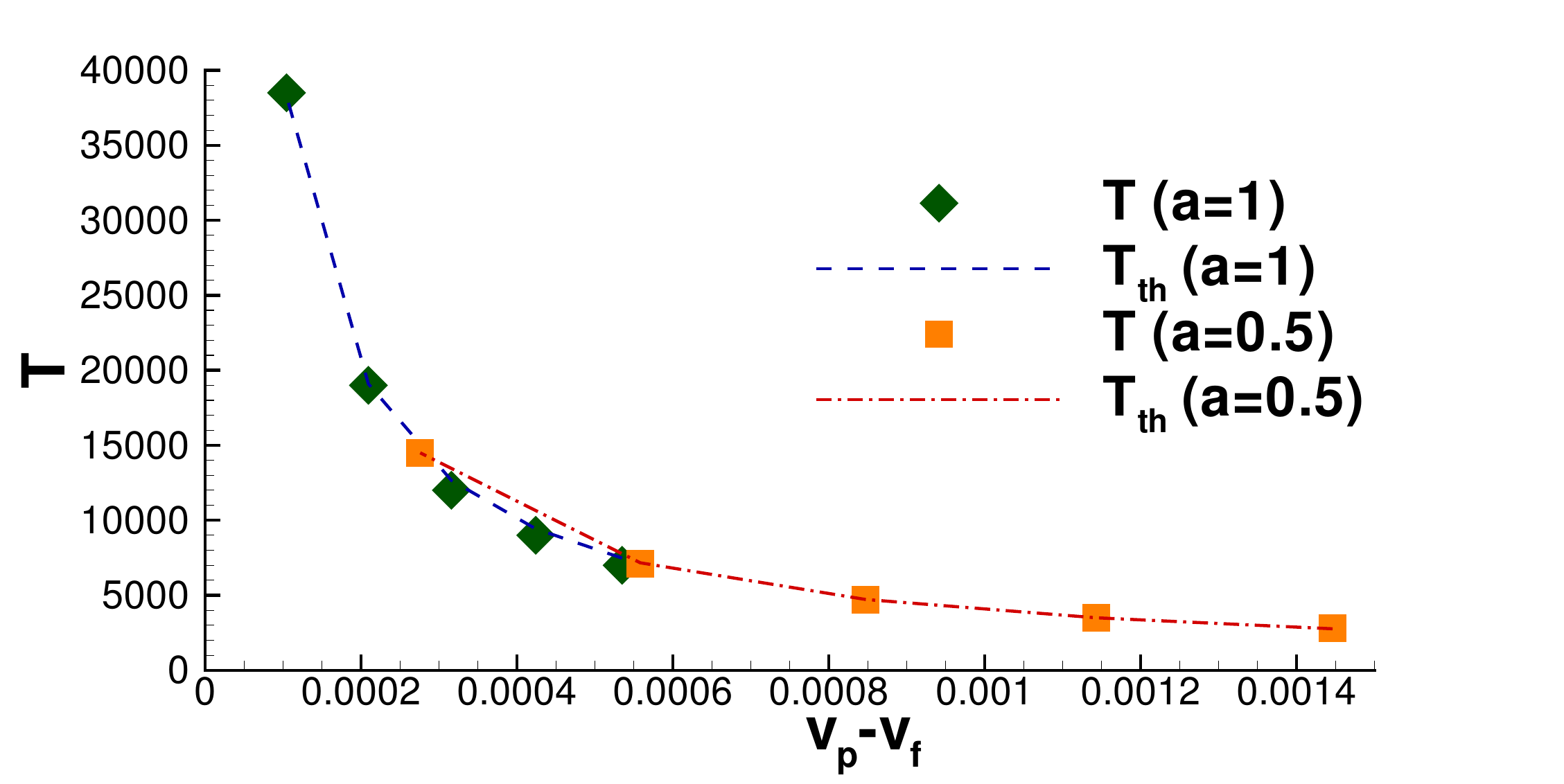}
\caption{Top panel: trajectory in the $x-y$ for the sphere of radius $a=1$
released at several wall normal initial positions 
$h_0$ which ranges from $0.5$ (straight line) to
$0.45$ (higher frequency curve). 
Inset: close up view of the trajectory at the initial stages.
Bottom panel: the fast oscillation period of $y[x(t)]$ (symbols) is compared 
with the model estimate (dashed line, eq.\ref{eq:oscit3}) as a function 
of the velocity difference $(v_p-v_f)$. 
}
\label{fig:f_tosci}
\end{figure}
%%%%%%%%%%%%%%%%%%%%%%%%%%%%%%%%%%%%%%%%%%%%%%%%%%%%%%%%%%%%%%%%%%%%%%%%%%%%

In the top panel of Fig. \ref{fig:f_tosci} different $xy$ trajectories are 
reported for  $a=1$,  $B=1$,  different initial positions $(0,0,h_0)$ and
$T=114$. The trajectory is characterized by two different wave-lengths. 
The shortest one is not appreciated on the scale of the plot, 
see the close-up in the inset. The long wavelength behavior strongly depends on 
the initial gap $h_0$ and
dominates the large scale motion. Its spatial period progressively increases as 
$h_0$  approaches the critical value $h_c= 0.5$ where the periodicity
is lost (infinite oscillation period) and the particle drifts steadily. 
In this limiting conditions the trajectory degenerates in a rectilinear motion with 
superimposed the aforementioned small scale oscillations. 
The observed behavior can be understood in terms of a simplified model amenable of
analytical solution. The model adopts a few assumptions that are 
reasonably well justified by the results described in the previous sections. 
A first simplification consists in freezing the $h$ dependence 
of the mobility coefficient $M_{22}(x,h) \simeq M_{22}(x,h_0)$. Successively, the 
$x$ dependence is approximated by its dominating Fourier modes 
$M_{22}(x,h_0) \simeq \overline{M}_{22} + b_1 \sin\left(\frac{2\pi}{\lambda} x\right)$,
where $\overline{M}_{22}$ and $b_1$ are mean value and first harmonic amplitude, 
respectively. Finally the motion along the $x$ direction under the action of 
the constant force $F_1$ is roughly approximated as $x(t) = v_p t + x_0$, 
where $v_p$ is the mean cross-stripe particle velocity. Equation ~(\ref{eq:oscit1})
than becomes
%%%%%%%%%%%%%%%%%%%%%%%%%%%%%%%%%%%%%%%%%%%%%%%%%%%%%%%%%%%%%%%%%%%%%%%%%%%%
\begin{equation}
\frac{dy}{dt} =
\left\{ \overline{M}_{22} + b_1 \sin\left[\frac{2\pi}{\lambda}  \left(v_pt + x_0 \right) \right]\right\}
 B \sin\left(\frac{2\pi t}{T}\right)  \ .
\label{eq:oscit2}
\end{equation}
%%%%%%%%%%%%%%%%%%%%%%%%%%%%%%%%%%%%%%%%%%%%%%%%%%%%%%%%%%%%%%%%%%%%%%%%%%%%
The solution follows as
%%%%%%%%%%%%%%%%%%%%%%%%%%%%%%%%%%%%%%%%%%%%%%%%%%%%%%%%%%%%%%%%%%%%%%%%%%%%
\begin{eqnarray}
y(t) & = & 
\label{eq:oscit3}
\frac{B \lambda b_1}{2\pi(v_p-v_f)}\sin\left(\frac{2\pi}{\lambda}(v_p-v_f)t + \phi \right) \\ 
& - & 
\label{eq:oscit4}
\frac{B \lambda b_1}{2\pi(v_p+v_f)}\sin\left(\frac{2\pi}{\lambda}(v_p+v_f)t+\phi \right) \\ 
& - & 
\label{eq:oscit5}
\frac{B \overline{M}_{22} \lambda}{2\pi v_f}\cos\left(\frac{2\pi}{\lambda}v_ft \right)  
+ y_0
\end{eqnarray}
%%%%%%%%%%%%%%%%%%%%%%%%%%%%%%%%%%%%%%%%%%%%%%%%%%%%%%%%%%%%%%%%%%%%%%%%%%%%
where $\phi = 2\pi x_0 /\lambda$ is the initial phase. In the above expression 
$v_f = \lambda/T$ is a characteristic velocity of the system fixed by the wall 
pattern spacing $\lambda$ and the forcing  period $T$. The time dependence of 
$y(t)$ results from the superimposition of three signals with different frequencies.  
The fundamental frequency is determined by the forcing period, term (\ref{eq:oscit5}), 
and the corresponding amplitude is proportional to the transverse mean mobility  
$\overline{M}_{22}$. This fundamental contribution remains the only 
relevant one far from the wall, where the $x$-dependence of $M_{22}$ becomes 
negligible (vanishing $b_1$), i.e.  the surface effectively behaves as a homogeneous 
wall.  The two sideband contributions, (\ref{eq:oscit3}) and (\ref{eq:oscit4}), 
follow from the interaction between the forcing frequency and the 
surface pattern, as shown by their proportionality to the transverse forcing 
amplitude $B$ and to the first Fourier coefficient of the pattern $b_1$.
For particles advancing in the positive $x$-direction ($v_p > 0$), the first 
side-band term (\ref{eq:oscit3}) is the more interesting one, since by proper 
tuning the relative velocity $v_p - v_f$ can be made to vanish.
The other side-band term, (\ref{eq:oscit4}), is typically irrelevant since its 
amplitude is too small to be detected in comparison with the fundamental 
contribution given that, in general, $\overline{M}_{22} \gg b_1$. 
For this reason such small oscillations can not be observed in Fig.~\ref{fig:f_tosci}. 
Concerning the first side-band term, in the limit $v_p - v_f \to 0$ its amplitude 
diverges while at the same time the frequency $2 \pi (v_p-v_f) /\lambda$  
vanishes, yielding the limiting behavior

%%%%%%%%%%%%%%%%%%%%%%%%%%%%%%%%%%%%%%%%%%%%%%%%%%%%%%%%%%%%%%%%%%%%%%%%%%%%
\begin{equation}
\label{eq:oscit6}
\lim_{v_p - v_f \to 0} y(t) = \frac{1}{2} B b_1 t \, .
\end{equation}
%%%%%%%%%%%%%%%%%%%%%%%%%%%%%%%%%%%%%%%%%%%%%%%%%%%%%%%%%%%%%%%%%%%%%%%%%%%%
\textcolor{black}{
This term, combined with $x(t) = v_p t + x_0$, gives reason of the (average) 
rectilinear trajectory of Fig.~\ref{fig:f_tosci} that occurs
when the initial condition $h_0$, $x_0$ selects a mean particle velocity $v_p(h_0,x_0)$ matching 
the system characteristic velocity $v_f$.}
Such case corresponds to resonance, since the particle exactly travels one wall 
pattern wave-length $\lambda$ in one oscillation period $T$ of the transversal
force. In these conditions a strong amplification of the particle transversal motion 
occurs. To complete the discussion, the bottom panel of Fig.~\ref{fig:f_tosci} 
shows the period of the fast oscillations of the in-plane trajectory $y[x(t)]$ as 
a function of the velocity difference $(v_p-v_f)$. Also reported are the 
corresponding values  from the model system,  term (\ref{eq:oscit3}). The 
comparison confirms the overall effectiveness of the simplified model.

\section{Conclusion}

The paper exploited the potential of the Boundary Element Method (BEM)
in the context of creeping flows in the Stokes regime. In a number of
applications at the micro-scales the fluid boundary condition at solid walls
is best described by a perfect-slip or partial slip boundary condition. The BEM  
proved successful in dealing with such complex surfaces characterized by a 
combination of perfect-slip (PS) and no-slip (NS) regions. In the geometry 
addressed in the present paper, the PS regions on the wall form a 
parallel-striped pattern able to effectively  model an actual superhydrophobic surface where gas bubbles are trapped in parallel grooves. 

In this context the hydrodynamics of a spherical particle moving 
close to the patterned surface has been characterized in terms of the mobility 
tensor. The wall pattern induces a characteristic spatial behavior of the 
mobility. Different regions have been identified. As expected, near the wall the 
mobility is larger in correspondence of the PS regions than in correspondence of 
the NS regions. In contrast, in the far field this behavior is reversed, i.e. 
the mobility becomes (slightly) larger when the sphere is just above a NS region. 
The modulation effects induced by the patterning become 
progressively weaker increasing the distance from the wall. In the far
field the oscillation in $M_{11}$ is less than $5\%$ of the average value 
$\overline{M}_{11}$. The wall normal distance $h_{inv}$ that separates the two 
regimes decreases with the sphere radius. Interestingly, irrespectively of 
the particle radius, the average mobility $\overline{M}_{11}$ for a given gap 
$h$ is well described by an effective model consisting of a homogeneous wall 
equipped with a partial slip boundary condition accounting for an 
effective slip-length, see e.g~\cite{philip1972flows,ng2010apparent}. 
The present results provide solid physical ground for the safe application 
of analytical models based on homogenization techniques~\cite{asmolov2011drag}.
This behavior is particularly evident for large particles 
(diameter larger than the stripe dimension) where the far field regime
sets in very close to the wall. It follows that, with the exception of a very 
thin area ($h < h_{inv}$), the mobility field is well approximated by the 
effective wall model. 

The characteristic spatial dependence of the mobility matrix can be potentially 
exploited for selective particle separation, as confirmed by simple numerical 
experiments. For instance, towing the sphere with a constant force in a wall 
parallel plane leads to a deviation of the trajectory.
The resulting drift angle is in agreement with results discussed 
in \cite{zhang2012separation} in the context of molecular dynamics 
simulations. Larger drift angles can be achieved by  other kind of
forcing, as in the case of a spatially or temporally oscillating force parallel 
to the stripes. In the second case in particular, a strong resonant effect 
occurs when the stripe-crossing frequency matches the external forcing 
frequency. This effect allows in principle to separate given particles from 
their neighbors.

As a final comment it is worth stressing that the model here described for the 
superhydrophobic surfaces relies on the assumption of a flat liquid-gas 
meniscus (PS region). A number of studies show that this is hardly strictly 
true also in simple systems
\cite{steinberger2007high,giacomello2012cassie,giacomello2012metastable,bolognesi2012}.
However, the present results show that a detailed  description of the surface 
pattern is not required unless the particle is very close to the wall. Indeed 
it was shown that it is actually often sufficient to model the surface as a 
homogeneous flat wall with a suitable apparent slip length tensor. In this 
effective description the curvatures of the menisci enters the problem by 
affecting the apparent slip-length, 
see e.g.\cite{teo2010flow,sbragaglia2007note,gentili2013water}. 
Still it should not be overlooked that the near wall behavior is dominated by 
the local effects emphasizing the role of the actual shape of the liquid/gas 
interface. The BEM approach here used can easily be extended to tackle such 
conditions to deal with  composite boundaries
on which perfect-slip, no-slip and partial slip boundary conditions are imposed.
This opens the way to the characterization of more complex systems such as 
slipping Janus particles motion~\cite{boymelgreen2011theoretical} and  
micro-swimmers~\cite{shum2012effects,zhu2013low}.

\appendix

\section{Boundary integral formulation} \label{sec:BEM}
%%%%%%%%%%%%%%%%%%%%%%%%%%%%%%%%%%%%%%%%%%%%%%%%%%%%%%%%%%%%%%%%%%%%%%%%%%%%
In this section the application of the Boundary Element Method (BEM) to the 
present geometrical configuration is shortly described, 
see e.g.~\cite{pozrikidis1992boundary,kim2005microhydrodynamics} for
further details.
 
Due to linearity, the constant coefficient Stokes problem, eq.~(\ref{eq:stokes}), 
can be recast into a boundary integral representation formula
%%%%%%%%%%%%%%%%%%%%%%%%%%%%%%%%%%%%%%%%%%%%%%%%%%%%%%%%%%%%%%%%%%%%%%%%%%%%
\begin{eqnarray}
\label{eq:integral_eq2}
E(\vxi) u_j(\vxi)& = &\frac{1}{8 \pi}
\int_{\partial B}  t_i(\vx) G_{ij}(\vx,\vxi) \, dS_{B}  \nonumber  \\
&-&\frac{1}{8 \pi } \int_{W_{NS}}  t_i(\vx) G_{ij}(\vx,\vxi) \, dS_{NS}  \nonumber  \\
&-&\frac{1}{8 \pi } \int_{ W_{PS}}  t_i(\vx) G_{ij}(\vx,\vxi) \, dS_{PS}  \nonumber \\
&-& \frac{1}{8\pi} \int_{\partial B} u_i(\vx)  \CT_{ijk} (\vx,\vxi) n_k(\vx) \, dS_{B} \nonumber \\
&-& \frac{1}{8\pi} \int_{W_{NS}} u_i(\vx)  \CT_{ijk} (\vx,\vxi) n_k(\vx) \, dS_{NS} \nonumber  \\
&-& \frac{1}{8\pi} \int_{W_{PS}} u_i(\vx)  \CT_{ijk} (\vx,\vxi) n_k(\vx) \, dS_{PS}.
\end{eqnarray}
%%%%%%%%%%%%%%%%%%%%%%%%%%%%%%%%%%%%%%%%%%%%%%%%%%%%%%%%%%%%%%%%%%%%%%%%%%%%
In equation (\ref{eq:integral_eq}) the boundary $\partial \Omega$ of the fluid domain
is explicitly decomposed in three parts: the particle surface $\partial B$, 
the no-slip stripes on the patterned wall, collectively denoted  $W_{NS}$, and the 
complementary part of the wall with the perfect slip  stripes $W_{PS}$.  
\textcolor{black}{The contributions arising from the portion of the boundary at infinity 
(not included in the present formulation where the fluid is assumed to be at rest) 
can be easily incorporated under suitable assumption on the asymptotic behavior of the field.
The effects of body forces like gravity or electric fields can be 
accounted for by a convolution integral extended to the fluid domain 
between the force and the free space Green's tensor (see below).}
In representation (\ref{eq:integral_eq2}), $t_i$ and $u_i$, $i=1,\ldots,3$, 
are the Cartesian components of the surface stress and velocity, respectively. 
The free  space Green's function (the so-called steady Stokeslet) is defined as
%%%%%%%%%%%%%%%%%%%%%%%%%%%%%%%%%%%%%%%%%%%%%%%%%%%%%%%%%%%%%%%%%%%%%%%%%%%%
\begin{equation}
G_{ij}(\vr)=\left(\frac{\delta_{ij}}{r} + \frac{r_i r_j}{r^3}\right)
\label{eq:Green} \, ,
\end{equation}
%%%%%%%%%%%%%%%%%%%%%%%%%%%%%%%%%%%%%%%%%%%%%%%%%%%%%%%%%%%%%%%%%%%%%%%%%%%%
and the associated stress tensor is 
%%%%%%%%%%%%%%%%%%%%%%%%%%%%%%%%%%%%%%%%%%%%%%%%%%%%%%%%%%%%%%%%%%%%%%%%%%%%
\begin{equation}
\CT_{ijk}(\vr) =-6 \frac{r_i r_j r_k }{r^5}. 
\label{eq:Green_stress}
\end{equation}
%%%%%%%%%%%%%%%%%%%%%%%%%%%%%%%%%%%%%%%%%%%%%%%%%%%%%%%%%%%%%%%%%%%%%%%%%%%%
In the above expression, $G_{ij}(\vr)$ provides the contribution to the j-th 
velocity component at $\vxi$ due to a concentrated force acting in the i-th 
direction at $\vx$. The associated Green's stress tensor, as always, should be 
contracted with the outward unit normal $n_k(\vx)$  to the boundary $\partial \Omega$  in order to provide the 
effect on the j-th velocity component at $\vxi$  of the i-th boundary velocity at 
$\vx$. The vector $\vr$ is defined as $\vr=\vx-\vxi$ with $r=\sqrt{r_k r_k}$ its 
modulus. In eq.~(\ref{eq:integral_eq}) $E(\vxi)=1$ when $\vxi\in\Omega$ 
and $E(\vxi)=1/2$ for $\vxi\in\partial\Omega$ (the existence of a regular tangent 
plane is assumed throughout). When $\vxi\in\partial\Omega$, representation~(\ref{eq:integral_eq}) 
becomes a boundary integral equation where the unknowns can either be the three 
stress vector components $t_i$, the three velocity components $u_i$, or a 
combination thereof, depending on the boundary conditions assigned on the specific 
portion of boundary. This approach allows to discretize only the boundary surfaces of 
the flow domain instead of considering the entire volume. This results in: 1)
a substantial reduction of the number of unknowns; 2) the possibility to easily 
specify different kinds of boundary condition on different surface patches; 3) the 
simple update of the geometry when dealing with time dependent configurations. Once 
the boundary $\partial\Omega$ is discretized into panels, eq.~(\ref{eq:integral_eq}) 
is recast into an algebraic linear system whose solution can be achieved by 
standard linear algebra packages. In simple cases, part of the boundary can be 
accounted for by symmetry, like it happens for a flat homogeneous wall.

In this paper, given the generality of the boundary condition to be used at the 
wall (either patterned perfect/no-slip stripes or effective slip Navier-like 
boundary conditions) the complete formulation of the boundary integral problem based 
on the free-space Green's function has been retained, with the use of the wall 
Green's function demanded of providing reference results for accuracy tests.

Finally, a few more words may be useful as concerning the specific boundary 
conditions used in the paper. On the perfect-slip boundary patches, $W_{PS}$, the 
normal velocity vanishes $u_{\perp}=0$ due to impermeability while  the tangential 
velocity $\vu_{\parallel}$ (two Cartesian components) is unknown. Moreover, the 
tangential stress ${\bf t}_{\parallel}$ vanishes by perfect slip such that the 
stress is aligned to the normal, $t_i=-\Phi(\vx) n_i$, with $\Phi$ representing a 
further scalar  unknown.  On the no-slip surfaces, $W_{NS}$ and $\partial B$, 
velocities are completely assigned  while stresses are unknown. 
Concerning the partial slip condition used as an effective model of the stripe 
pattern, zero normal velocity $u_{\perp}=0$ at the wall are implied, while tangential 
velocities and stresses are coupled by the Navier condition \cite{lauga2007handbook},
%%%%%%%%%%%%%%%%%%%%%%%%%%%%%%%%%%%%%%%%%%%%%%%
\begin{eqnarray}
\vu_{\parallel} = \vec{\ell_s}  \vec{n} \cdot ( \grad{\vu} + (\grad{\vu})^T)
\cdot (\vec{1}-\vec{n} \otimes \vec{n}) \ .
\label{eq:navier}
\end{eqnarray}
%%%%%%%%%%%%%%%%%%%%%%%%%%%%%%%%%%%%%%%%%%%%%%%
Here $\vec{\ell_s}$ is a $2\times2$ symmetric~\cite{kamrin2010effective} tensor 
describing the directionally dependent slip-length. In the present case this 
tensor is diagonalized when expressed in stripe-parallel and stripe-normal  
Cartesian coordinates. In this case the two diagonal entries are different as a 
consequence of the orientation of the stripe pattern. For a flat wall  
eq.~(\ref{eq:navier}) is rewritten as 
%%%%%%%%%%%%%%%%%%%%%%%%%%%%%%%%%%%%%%%%%%%%%%%
\begin{equation}
\vec{u}_{\parallel}=\vec{\ell_s} \cdot \vec{t_{\parallel}} \, ,
\label{eq:our_navier}
\end{equation}
%%%%%%%%%%%%%%%%%%%%%%%%%%%%%%%%%%%%%%%%%%%%%%%
that is the vectorial form of eq.
(\ref{eqn:Navier_BC}), the two non-zero components of $\vec{\ell_s}$ being 
reported in eq. \ref{eqn:slip_lenghts}~\cite{philip1972flows,ng2010apparent}.
The boundary integral equation (\ref{eq:integral_eq}) supplemented with  
eq.~(\ref{eq:our_navier}) provides a closed system that after inversion determines 
all the unknowns involved in the problem.
%%%%%%%%%%%%%%%%%%%%%%%%%%%%%%%%%%%%%%%%%%%%%%%%%%%%%%%%%%%%%%%%%%%%%%%%%%%%
%%%%%%%%%%%%%%%%%%%%%%%%%%%%%%%%%%%%%%%%%%%%%%%%%%%%%%%%%%%%%%%%%%%%%%%%%%%%
%%%%%%%%%%%%%%%%%%%%%%%%%%%%%%%%%%%%%%%%%%%%%%%%%%%%%%%%%%%%%%%%%%%%%%%%%%%%
%%%%%%%%%%%%%%%%%%%%%%%%%%%%%%%%%%%%%%%%%%%%%%%%%%%%%%%%%%%%%%%%%%%%%%%%%%%%
%%%%%%%%%%%%%%%%%%%%%%%%%%%%%%%%%%%%%%%%%%%%%%%%%%%%%%%%%%%%%%%%%%%%%%%%%%%%
%%%%%%%%%%%%%%%%%%%%%%%%%%%%%%%%%%%%%%%%%%%%%%%%%%%%%%%%%%%%%%%%%%%%%%%%%%%%
\bibliographystyle{plain}       % APS-like style for physics
\bibliography{pimponi_microflu2013}   % name your BibTeX data base

\begin{thebibliography}{10}
\providecommand{\url}[1]{{#1}}
\providecommand{\urlprefix}{URL }
\expandafter\ifx\csname urlstyle\endcsname\relax
  \providecommand{\doi}[1]{DOI \discretionary{}{}{}#1}\else
  \providecommand{\doi}{DOI \discretionary{}{}{}\begingroup
  \urlstyle{rm}\Url}\fi

\bibitem{nosonovsky2009superhydrophobic}
M.~Nosonovsky, B.~Bhushan, Current Opinion in Colloid \& Interface Science
  \textbf{14}(4), 270 (2009)

\bibitem{bottiglione2012role}
F.~Bottiglione, G.~Carbone, Langmuir \textbf{29}(2), 599 (2012)

\bibitem{ybert2007achieving}
C.~Ybert, C.~Barentin, C.~Cottin-Bizonne, P.~Joseph, L.~Bocquet, Physics of
  fluids \textbf{19}, 123601 (2007)

\bibitem{ng2010apparent}
C.~Ng, C.~Wang, Microfluidics and Nanofluidics \textbf{8}(3), 361 (2010)

\bibitem{lee2011influence}
C.~Lee, C.~Kim, Langmuir: the ACS journal of surfaces and colloids
  \textbf{27}(7), 4243 (2011)

\bibitem{vinogradova2011wetting}
O.I. Vinogradova, A.V. Belyaev, Journal of Physics: Condensed Matter
  \textbf{23}(18), 184104 (2011)

\bibitem{philip1972flows}
J.~Philip, Zeitschrift f{\"u}r Angewandte Mathematik und Physik (ZAMP)
  \textbf{23}(3), 353 (1972)

\bibitem{li2009critical}
Z.~Li, Physical Review E \textbf{80}(6), 061204 (2009)

\bibitem{chinappi2008mass}
M.~Chinappi, S.~Melchionna, C.M. Casciola, S.~Succi, The Journal of Chemical
  Physics \textbf{129}, 124717 (2008)

\bibitem{benzi2006mesoscopic}
R.~Benzi, L.~Biferale, M.~Sbragaglia, S.~Succi, F.~Toschi, EPL (Europhysics
  Letters) \textbf{74}, 651 (2006)

\bibitem{cottin2004dynamics}
C.~Cottin-Bizonne, C.~Barentin, {\'E}.~Charlaix, L.~Bocquet, J.~Barrat, The
  European Physical Journal E: Soft Matter and Biological Physics
  \textbf{15}(4), 427 (2004)

\bibitem{gentili2013water}
D.~Gentili, M.~Chinappi, G.~Bolognesi, A.~Giacomello, C.~Casciola, Meccanica
  pp. 1--9 (2013)

\bibitem{chinappi2011tilting}
M.~Chinappi, F.~Gala, G.~Zollo, C.M. Casciola, Philosophical Transactions of
  the Royal Society A: Mathematical, Physical and Engineering Sciences
  \textbf{369}(1945), 2537 (2011)

\bibitem{chinappi2010intrinsic}
M.~Chinappi, C.M. Casciola, Physics of Fluids \textbf{22}, 042003 (2010)

\bibitem{huang2008water}
D.~Huang, C.~Sendner, D.~Horinek, R.~Netz, L.~Bocquet, Physical review letters
  \textbf{101}(22), 226101 (2008)

\bibitem{zhang2012molecular}
H.~Zhang, Z.~Zhang, H.~Ye, Microfluidics and Nanofluidics pp. 1--9 (2012)

\bibitem{zhu2012reconciling}
L.~Zhu, C.~Neto, P.~Attard, Langmuir \textbf{28}(20), 7768 (2012)

\bibitem{cottin2008nanohydrodynamics}
C.~Cottin-Bizonne, A.~Steinberger, B.~Cross, O.~Raccurt, E.~Charlaix, Langmuir
  \textbf{24}(4), 1165 (2008)

\bibitem{pan2012role}
Y.~Pan, B.~Bhushan, Journal of Colloid and Interface Science  (2012)

\bibitem{vinogradova1995drainage}
O.I. Vinogradova, Langmuir \textbf{11}(6), 2213 (1995)

\bibitem{belyaev2010effective}
A.V. Belyaev, O.I. Vinogradova, Journal of Fluid Mechanics \textbf{652}, 489
  (2010)

\bibitem{happel1965low}
J.R. Happel, H.~Brenner, \emph{Low Reynolds number hydrodynamics: with special
  applications to particulate media}, vol.~1 (Springer, 1965)

\bibitem{kim2005microhydrodynamics}
S.~Kim, S.~Karrila, \emph{Microhydrodynamics: principles and selected
  applications} (Dover Publications, 2005)

\bibitem{pozrikidis1992boundary}
C.~Pozrikidis, \emph{Boundary integral and singularity methods for linearized
  viscous flow}.
\newblock 8 (Cambridge University Press, 1992)

\bibitem{brady1988stokesian}
J.F. Brady, G.~Bossis, Annual review of fluid mechanics \textbf{20}, 111 (1988)

\bibitem{pozrikidis2002practical}
C.~Pozrikidis, \emph{A practical guide to boundary element methods with the
  software library BEMLIB} (CRC, 2002)

\bibitem{blake1971note}
J.~Blake, in \emph{Proc. Camb. Phil. Soc}, vol.~70 (Cambridge Univ Press,
  1971), vol.~70, pp. 303--310

\bibitem{landau1987fluid}
L.~Landau, \emph{Fluid Mechanics: Volume 6 (Course Of Theoretical Physics)
  Author: LD Landau, EM Lifshitz, Publisher: Bu} (Butterworth-Heinemann, 1987)

\bibitem{goldman1967slow}
A.~Goldman, R.G. Cox, H.~Brenner, Chemical Engineering Science \textbf{22}(4),
  637 (1967)

\bibitem{batchelor1976brownian}
G.~Batchelor, Journal of Fluid Mechanics \textbf{74}(01), 1 (1976)

\bibitem{jeffrey1984calculation}
D.~Jeffrey, Y.~Onishi, Journal of Fluid Mechanics \textbf{139}(1), 261 (1984)

\bibitem{zhang2012separation}
R.~Zhang, J.~Koplik, Physical Review E \textbf{85}(2), 026314 (2012)

\bibitem{zhou:194706}
J.~Zhou, A.V. Belyaev, F.~Schmid, O.I. Vinogradova, The Journal of Chemical
  Physics \textbf{136}(19), 194706 (2012).
\newblock \doi{10.1063/1.4718834}.
\newblock \urlprefix\url{http://link.aip.org/link/?JCP/136/194706/1}

\bibitem{asmolov2011drag}
E.S. Asmolov, A.V. Belyaev, O.I. Vinogradova, Physical Review E \textbf{84}(2),
  026330 (2011)

\bibitem{steinberger2007high}
A.~Steinberger, C.~Cottin-Bizonne, P.~Kleimann, E.~Charlaix, Nature materials
  \textbf{6}(9), 665 (2007)

\bibitem{giacomello2012cassie}
A.~Giacomello, S.~Meloni, M.~Chinappi, C.M. Casciola, Langmuir \textbf{28}(29),
  10764 (2012)

\bibitem{giacomello2012metastable}
A.~Giacomello, M.~Chinappi, S.~Meloni, C.M. Casciola, Physical Review Letters
  \textbf{109}(22), 226102 (2012)

\bibitem{bolognesi2012}
G.~Bolognesi, C.~Pirat, E.~Cottin-Bizonne, M.~Guene, J.~Teisseire, Soft Matter
  (Under Review)

\bibitem{teo2010flow}
C.~Teo, B.~Khoo, Microfluidics and Nanofluidics \textbf{9}(2), 499 (2010)

\bibitem{sbragaglia2007note}
M.~Sbragaglia, A.~Prosperetti, Physics of Fluids \textbf{19}, 043603 (2007)

\bibitem{boymelgreen2011theoretical}
A.M. Boymelgreen, T.~Miloh, Physics of Fluids \textbf{23}, 072007 (2011)

\bibitem{shum2012effects}
H.~Shum, E.~Gaffney, Physics of Fluids \textbf{24}, 061901 (2012)

\bibitem{zhu2013low}
L.~Zhu, E.~Lauga, L.~Brandt, Journal of Fluid Mechanics \textbf{726} (2013)

\bibitem{lauga2007handbook}
E.~Lauga, M.~Brenner, H.~Stone.
\newblock Handbook of experimental fluid mechanics (2007)

\bibitem{kamrin2010effective}
K.~Kamrin, M.Z. Bazant, H.A. Stone, Journal of Fluid Mechanics \textbf{658},
  409 (2010)

\end{thebibliography}

\end{document}